\documentclass[journal]{IEEEtran}
\ifCLASSINFOpdf
\else
\fi
 \usepackage{pdfsync}

\usepackage{multicol,color}
\usepackage{ragged2e}
\usepackage{amssymb}
\usepackage{amsmath}
\usepackage{tabularx}
\usepackage{booktabs}

\newtheorem{theorem}{\textbf{Theorem}}

\newtheorem{lemma}{\textbf{Lemma}}
\newtheorem{remark}{\textbf{Remark}}
\newtheorem{proposition}{\textbf{Proposition}}

\usepackage{graphics}
\usepackage{algorithm}
\usepackage{algorithmicx}
\usepackage{algpseudocode}
\usepackage{subfigure}

\usepackage[]{hyperref} 

\hypersetup{
	colorlinks=true,
	linkcolor=blue,
	filecolor=blue,      
	urlcolor=blue,
	citecolor=cyan,
}

\usepackage{array}
\usepackage{url}

\usepackage{graphicx}

\makeatletter

\renewcommand{\maketag@@@}[1]{\hbox{\m@th\normalsize\normalfont#1}}%

\makeatother
\newtheorem{assumption}{\textbf{Assumption}}
\begin{document}
%
\title{  
	Multi-Cluster Aggregative Games: A Linearly
	Convergent Nash Equilibrium Seeking Algorithm \\and its Applications in Energy Management}
%
%
%

\author{
	Yue Chen and Peng Yi

\thanks{The authors are with the Department of Control Science and Engineering,
	Tongji University, Shanghai, 201804, China, and also with  Shanghai Institute of Intelligent Science and Technology, Tongji University, Shanghai 200092, China. 
	Corresponding author: Peng Yi.
	(email: chenyue\_j@tongji.edu.cn, yipeng@tongji.edu.cn).  }
}

%
%

\markboth{ }
{Shell \MakeLowercase{\textit{et al.}}: Bare Demo of IEEEtran.cls for IEEE Journals}
%




\maketitle
\begin{abstract}
We propose a type of non-cooperative game, termed \textit{multi-cluster aggregative game}, which is composed of clusters as players, where each cluster consists of collaborative agents with cost functions depending on their own decisions and \textit{the aggregate quantity} of each participant cluster to modeling large-scale and hierarchical multi-agent systems. 
This novel game model is motivated by decision-making problems in competitive-cooperative network systems with large-scale nodes, such as \textit{the Energy Internet}.
To address challenges arising in seeking Nash equilibrium for such network systems,
we develop an algorithm with a hierarchical communication topology 
which is a hybrid with distributed and semi-decentralized
protocols.
The upper level consists of
cluster coordinators estimating the aggregate quantities with
local communications, while the lower level is 
cluster subnets composed of its coordinator and agents
aiming to track the gradient of the corresponding cluster. 
In particular, the clusters exchange the aggregate quantities instead
of their decisions to relieve the burden of communication.
Under strongly monotone and mildly Lipschitz continuous assumptions, we rigorously prove that the algorithm linearly converges to a Nash equilibrium with a fixed step size.
We present the applications in the context of the Energy Internet.
Furthermore, the numerical results verify the effectiveness of the algorithm.
\end{abstract}


%
\IEEEpeerreviewmaketitle

\section{Introduction}
\par
Game theory is widely employed in modeling multi-agent systems, while Nash equilibrium serves as a common tool to characterize a stable state where no player unilaterally changes their decision.
However, as the need for modeling large-scale complex systems and tasks increases, new challenges have emerged in seeking  Nash equilibrium \cite{yi2022survey}. 
One of these challenges is dealing with hierarchical properties, including both competition and collaboration \cite{wang2022cooperative}. 
These properties exist in many real-world applications, such as health-care networks \cite{peng2009coexistence},
task allocation networks \cite{shehory1998methods},
and electricity markets \cite{zeng2019}.
Moreover, mitigating the computational complexity brought by a large population size is a critical issue. 
In such multi-agent systems, agents usually lack full-decision information about all others, which requires an effective estimation and communication scheme \cite{pavel2019distributed, ma2019consensus}. 
This work is motivated by these problems and aims to design an appropriate game model for hierarchical multi-agent systems with a large population and propose an effective algorithm for seeking the Nash equilibrium.
\par
We introduce the following two game types, known as multi-cluster games and aggregative games, that pave a way to model an appropriate model for
large-population multi-agent systems with  hierarchical properties.
\subsection{Multi-cluster Games and Aggregative games}
The multi-cluster game is a significant game type that comprises multiple clusters (or coalitions \cite{ye2018nash}, \cite{zeng2017distributed}), where clusters are treated as players instead of agents, while agents in the same cluster have a common goal to minimize the cost of corresponding cluster. 
Although the virtual players are clusters, the real decision makers are agents, which implies
the appearance of competition and cooperation simultaneously.
Noted that such a hierarchical property requires the multi-cluster game model to take
both inter-cluster and intra-cluster interactions into account, where inter-cluster is the mutual  behavior of agents within the cluster, and intra-cluster is between clusters.
In \cite{ye2018nash} , \cite{deng2021generalized}, and \cite{zimmermann2021solving}, algorithms require communication between clusters is carried out by the underlying network which consists of agents chosen in each cluster, while communications within clusters are proceeded by networks composed of corresponding cluster agents.
In  \cite{zeng2019} and  \cite{zhou2022distributed}, there are leaders in charge of the communication between clusters (inter-graph),
and other agents obtain information by communicating with leaders (intra-graph).
\par 
The aggregative game is a type of non-cooperative game where the objectives of agents depend on the aggregate quantity rather than individual interactions, 
with wide-ranging applications including
traffic or transmission \cite{alpcan2002cdma} networks,
smart grids \cite{chen2014autonomous, shakarami2022distributed},
charging management \cite{ma2016efficient},
and network congestion control \cite{barrera2014dynamic}.
Many research works have studied algorithms for solving the Nash equilibrium for aggregative games under diverse conditions, such as 
in \cite{lei2021distributed}, monotone property regimes were considered while others usually need strict or strong monotonicity.
in \cite{gadjov2020single}, an operator theory approach is utilized, while every player has coupling linear constraints.
and  \cite{zhang2019distributed} consider uncertain perturbed nonlinear players.
However, the game model structure examined in these literature works is predominantly single-layered, with algorithm design primarily accomplished within the framework of single-layer communication.
\subsection{Motivations}
\par Although both game types have been studied for a wide range of applications, there are limited investigations into problems involving both large-population size and hierarchical properties (cooperation and competition). 
For networked multi-cluster games, there are challenges in designing  an effective distributed Nash equilibrium seeking algorithm when the population size of each cluster is large-scale.
Since agents cannot have full-information decisions of others in networked games,  there are two main communication protocols (namely, fully distributed and leader-follower-based)
in the aforementioned multi-cluster games.
In fully distributed cases such as \cite{nian2021distributed}, each agent relies entirely on the information exchange from neighbors to achieve an agreement.
In leader-follower-based cases such as  \cite{zeng2019}, leaders take responsibility to communicate with both intra-and inter-clusters.
However, both  approaches bring about   a significant communication and computation burden on the system,  since agents have to estimate all others' decisions,
which makes the dimensions of the estimator expand with the increase of agents.
In addition, we note that   the aggregative structure  in the aforementioned aggregative games can greatly reduce communication complexity. However, the interactions are only between non-cooperative individuals, which makes it incapable of modeling hierarchical and complex multi-agent systems.
\par 
Motivated by challenges that arise in seeking Nash equilibrium in large-scale and hierarchical multi-agent systems, we first propose a novel game model, named multi-cluster aggregative game.
It combines hierarchical properties and the aggregative structure to model the decision-making procedure of a large-population multi-agent system that involves non-cooperative  clusters and collaborative agents simultaneously.
Such games are appropriate for many practical applications,
especially as an effective scheme for modeling an energy management system.
In the following, we will introduce an important type of energy management system, named the Energy Internet, as an application to further clarify this.
\subsection{The Energy Internet}
\par
The Energy Internet is a system with the capability to accommodate diverse kinds of energy, c.f. \cite{huang2010future}. 
This system relies on many energy interfaces, termed energy hub, to physically connect the energy supply 
and demand sides. 
To facilitate the seamless flow of energy, 
each energy hub has an energy port that enables a certain amount of energy exchange. 
Energy generation devices, including wind farms, photovoltaic stations, boilers, gas wells, and turbines, are directly linked to the energy networks, as well as energy storage devices such as supercapacitors and pumped hydro storage system. 
The energy hubs can aggregate the energy demands (input and output) of residents within a certain area, 
and they can be seen as prosumers.
\par
In an Energy Internet system, various regions (communities) act as non-cooperative participants, while energy hubs within each region collaborate to minimize the overall cost of the respective region.
Each energy hub is an independent decision maker, whose cost is not only dependent on its own decision but also dependent on all other region's aggregate demands.
Such a multi-agent system can be modeled as a multi-cluster aggregative game.
To fulfill the ``plug and play" needs  (see e.g. \cite{sun2019energy} ) of the Energy Internet,
we propose a novel hierarchical network scheme, in which each cluster is a semi-decentralized structure as in \cite{belgioioso2017semi}, and coordinators interact in a distributed structure to ensure that their estimates are in agreement.x`
\par In many scenarios, such as demand response,  agents' cost functions rely on the aggregate quantity, cf., \cite{kebriaei2021multipopulation}, \cite{belgioioso2017semi}, and \cite{hu2022distributed} rather than the decision of every agent.
In particular, energy demands are more capable of being simplified as an aggregate quantity.
Consequently,  the multi-cluster aggregative game proves to be an appropriate model for large-scale energy management systems.
\subsection{Contributions}
\par
The main contributions are listed as follows: 
\begin{enumerate}
	\item  We propose a novel multi-cluster aggregative game which is a non-cooperative game involving competitive clusters and collaborative agents, whereas the cost function of each agent is not only dependent on its own decision but also the aggregate quantity of every cluster. To solve this game, we have to overcome two challenges. One is the partial-decision information regime, namely, each agent does not have all others' decisions, herein, the aggregate quantity of every cluster. The other one is the distributed information constraints, that is, agents do not have the aggregate gradient information of the corresponding cluster. 
	\item We propose a Nash equilibrium seeking algorithm with a linear convergence rate for a multi-cluster aggregative game under both partial-decision and distributed information constraints.
	The algorithm proceeds with the coordinations with both coordinators and agents, where coordinators exploit a consensus scheme to estimate the aggregate quantities and broadcast them, while agents receive the broadcast values, obtain the aggregate gradient by local tracking, and calculate Nash equilibrium via gradient descent.
	\\
	For the sake of relieving the communication burden of large-population clusters, we design a hierarchical communication topology involving distributed and semi-decentralized protocols. The topology between and within clusters is distributed, however, the interactions between coordinators and agents are semi-decentralized.\\
	The algorithm with such topology has the following features: one is the ``anonymity'' of the agent, namely, an agent only has to recognize its neighbors within the cluster rather than non-neighbor or other-cluster agents. 
	The other one is the ``lightweight'' feature, which means the algorithm only requires small calculation resources. 
	The last one is the ``plug and play'' feature, i.e., the requirements of adding a member coordinator or agent are easy to fulfill, which only needs the coordination of coordinators or agents.
	\item 
	We discuss the utilization of multi-cluster aggregative games in the context of an Energy Internet system by corresponding the aggregate and competitive-cooperative features to properties of the energy generation/consumption and the regions-hubs relationships in the Energy Internet.
	The numerical results verify the effectiveness of the algorithm.
\end{enumerate}
\par
The remainder of this paper is structured as follows.
We formulate a multi-cluster aggregative game and discuss its applications in the context of the Energy Internet
in section \ref{section2},  
introduce an algorithm with a hierarchical communication scheme in section \ref{section3}, 
present the convergence analysis in Section \ref{section4}, 
show the numerical simulation results in Section \ref{section5}, 
and give concluding remarks in Section \ref{section6}.
\subsection{Notation and Preliminaries}
We denote by $\mathbb{R}^m(\mathbb{R}^m_+) $ a $ m $-dimensional (non-negative) Euclidean space, 
where $m$ is a positive integer.
We utilize $ \mathcal{V}\triangleq\{1,2,\dots,m\} $ to denote a set contains integers from $1$ to $m$.
In this paper, all vectors are viewed as column vectors.
The vectors of $ m $-dimensional with all items being $1$ or $0$ are denoted by $\mathbf{1}_m$ or $\mathbf{0}_m$
while the identity matrix of $ m \times m $ is denoted as $ \boldsymbol{I}_m $.
Let $x$ and $ A $  be a vector and a matrix, and the transpose of $x$ and $ A $ is denoted by $ x^T $ and $ A^T $.
For simplicity, for all $ j\in \mathcal{V} $, $ col((x^j)_{ j \in \mathcal{V}}) $ is equivalent to $ [(x^1)^T, \dots, (x^m)^T ]^T $, where 
$ x^j $ is a vector. 
Similarity, when $ A^1, \dots, A^m  $ are matrices, $ diag((A^j)_{j\in \mathcal{V}}) $ refers to a block diagonal matrix with $A^1, \dots, A^m$ positioned on the diagonal.
For a function $ \theta(x^1, . . . , x^m) $, $ \nabla_{x^j} \theta $ 
denote
the gradient of $ \theta  $ with respect to $ x^j $.
\section{GAME FORMULATION} \label{section2}
\subsection{Multi-Cluster Aggregative Games}
Consider a multi-cluster aggregative game consisting of $ m $ clusters where each cluster is treated as a non-cooperative player. Every cluster
$ j\in  \mathcal{V} \triangleq \{1,2,\dots, m\} $ contains  $n_j $ number of agents. Then the number of agents in the game is  $ n\triangleq\sum_{j=1}^m n_j $.
Let $  x^j_i \in \mathbb{R}^q$ denote the decision of agent $ i  $ in cluster $ j $  where $ i \in \mathcal{S}_j\triangleq \{1,2,\dots, n_j\} $,
then  $ x^j=col((x^j_i)_{ i \in \mathcal{S}_j})\in \mathbb{R}^{n_jq}  $ is the   stacked strategy vector of cluster $ j $   piled up by $ x^j_i $.
Analogously,  $ \boldsymbol{x}\in \mathbb{R}^{nq} $  is defined as
$ \boldsymbol{x}=(x^1,\dots, x^j,\dots,x^m) $
, and $ \boldsymbol{x}^{-j} $ denotes the strategies of all clusters except cluster $ j $,
i.e.,
$ \boldsymbol{x}^{-j} \triangleq (x^1,\dots,x^{j-1},x^{j+1},\dots, x^m) $.
The strategy aggregate quantity of  cluster $ j $ is defined as follows
\[
v^j(x^j) \triangleq \frac{1}{n_j} \sum_{i=1}^{n_j}x^j_i, j\in \mathcal{V},
i\in \mathcal{S}_j.
\]
\par
The cost function of agent $ i $  in cluster $ j $ is given by
\begin{flalign}
	\theta^j_i(x^j_i,\boldsymbol{v}(\boldsymbol{x})) \triangleq \theta^j_i(x^j_i,v^1(x^1),\dots,v^m(x^m)),
\end{flalign}
where
 $ \boldsymbol{v}(\boldsymbol{x})\triangleq (v^1(x^1),\dots,v^m(x^m)):
\mathbb{R}^{n_1 q \times \dots \times n_m q} \rightarrow \mathbb{R}^{q\times \dots \times q}$,
and $ \theta^j_i: \mathbb{R}^q \times \mathbb{R}^{mq} \rightarrow \mathbb{R} $. Denote by
\small
\[
\boldsymbol{v}{(\boldsymbol{x}^{-j})}\triangleq(v^1{(x^1)},\dots,v^{j-1}(x^{j-1}),v^{j+1}({x^{j+1}}),\dots,v^m(x^m)),
\]
\normalsize
the aggregate vector except cluster $ j $.
 Each cluster's cost function is the sum of all agent's cost functions within that cluster, i.e., cluster $ j $'s cost function is given by
\begin{flalign} \label{ClusterFunc}
\theta^{j}\left(x^{j}, \boldsymbol{v}(\boldsymbol{x})\right) \triangleq
\sum_{i=1}^{n_{j}} \theta^{j}_i\left(x^{j}_i,\boldsymbol{v}(\boldsymbol{x})\right): \mathbb{R}^{n_j q}\times \mathbb{R}^{mq} \rightarrow \mathbb{R}.
\end{flalign}
\par For each cluster $ j\in \mathcal{V} $, the problem is given by
\begin{flalign} \label{cluster-agg}
 \operatornamewithlimits{minimize}_{x^j} & \quad \theta^{j}\left(x^{j}, \boldsymbol{v}(\boldsymbol{x})\right).
\end{flalign}
\par 
Note that the optimal problem (\ref{cluster-agg}) for each cluster, which involves the aggregate quantity of every cluster, is distinct from that in many of the existing research works.
This kind of games can be exploited in many multi-agent applications. 
We present an Energy Internet example in the following subsection.

\subsection{Example of The Energy Internet} \label{case2}
Consider an Energy Internet system that manages multiple regions, with each region having multiple energy hubs composing of an energy subnet, where each subnet corresponds to a cluster in the aforementioned game model. 
The energy subnets' optimal energy strategy decision problem can be recast into a Nash equilibrium seeking problem of a multi-cluster aggregative game.
\par The strategy vector of energy hub $ i $ in energy subnet $ j $ is
$ x^j_i\triangleq(x^j_{i,E},x^j_{i,H},x^j_{i,G}) $,
where $ x^j_{i,E} $, $ x^j_{i,H} $ and $ x^j_{i,G} $ are demand of electricity, heat and gas \cite{liang2019generalized}.
The positive and negative of the quantity represent the export and import behavior of the hub respectively.
The cost function of hub $ i $ is given by
$ \theta^j_i (x^j_i, \boldsymbol{v}(\boldsymbol{x}))  \triangleq c^j_i(x^j_i) -p^j(\boldsymbol{v}(\boldsymbol{x}))^Tx^j_i $,
where $ c^j_i(x^j_i) \triangleq \frac{1}{2} (x^j_i)^T Q^j_i x^j_i +(b^j_i)^T x^j_i + c^j_i $ stands for the cost of hub $ i $
with $ Q^j_i \in \mathbb{R}^{3\times3} $, $ b^j_i \in \mathbb{R}^{3} $, and $ c^j_i \in \mathbb{R} $ being private parameters of hub $ i $.
Herein,  $ p^j(\boldsymbol{v}(\boldsymbol{x})) \triangleq
d^j-C^j (v^j + a^{j} \sum_{l=1,l\neq j}^m v^l)   $
denotes the energy price vector of $j  \in \mathcal{V} $ that not only depends on its own
demand requirement, but also   on other clusters' aggregate quantities  $ v^l \triangleq \frac{1}{n_l} \sum_{i=1}^{n_l} x^l_i, \forall l\in \mathcal{N}^j,
i \in \mathcal{S}_j  $, where
$ d^j \in \mathbb{R}^3 $ and 
$ C^j \in \mathbb{R}^{3 \times 3} $ are hub $ j $'s private parameters, and $ a^{j} \in [0,1] $  represents how much the price of cluster $ j $ is affected by other clusters. The vector $ \textbf{a} \triangleq col((a^j)_{j\in \mathcal{V}})$ is called impact factor which indicates the different policies corresponding to  different regions,
while the diagonal matrix $ C^j \in \mathbb{R}^{3\times3} $ is the policies according to different kinds of energy, in this regime, i.e.,  electricity, heat, and gas. 
\par 
It is necessary to divide hubs into clusters to fulfill the energy management in an Energy Internet system,
where a cluster is a subnet composed of energy hubs in one region.
The coordination of all energy hubs' energy consumption and generation requires hierarchical aggregative interactions between clusters and within clusters, 
whereas the existing studies are not capable of dealing with it.  
All in all, a Nash equilibrium seeking problem of a multi-aggregative game is an appropriate model for the Energy Internet's optimal energy strategies.
\subsection{Basic Analysis}
\par 
We present the analysis and the design motivation of the proposed algorithm in this subsection, after imposing the following assumption of the participants' cost functions in the game. 
\begin{assumption} \label{assumption_func}
	For each agent $ i\in \mathcal{S}_j $ in cluster $ j\in \mathcal{V}$,
	the cost function $ \theta^j_i (\cdot, \boldsymbol{v}(\boldsymbol{x}^{-j})) $ is convex in $ \mathbb{R}^{ q}  $
	for every fixed $ \boldsymbol{x}^{-j} \in \mathbb{R}^{(n-n_j)q} $, and
	$ \theta^j_i (x^j_i, \boldsymbol{v}) $
	is continuously differentiable in $(x_i^j, \boldsymbol{v} ) \in \mathbb{R}^{q} \times \mathbb{R}^{mq}  $.
\end{assumption}
\par Under Assumption \ref{assumption_func}, for all $ j\in \mathcal{V} $,
the cost function
$ \theta^j(\cdot, \boldsymbol{v}(\boldsymbol{x}^{-j})) $
is convex in $ \mathbb{R}^{n_jq}  $
and
$ \theta^j(x^j, \boldsymbol{v}) $
is continuously differentiable in
$(x^j, \boldsymbol{v}) \in \mathbb{R}^{n_jq} \times \mathbb{R}^{mq}  $.
\par Let $ z^j \triangleq \frac{1}{n_j} \sum_{i=1}^{n_j} x^j_i $ be a component of $ z\triangleq(z^1,\dots,z^j,\dots, z^m) \in \mathbb{R}^{mq}   $.
The gradient of  $ \theta^j(x^j,\boldsymbol{v}(\boldsymbol{x})) $ with respect to $ x^j $ is given by
\small \begin{flalign} 	 \label{optimalcondition}
	&\nabla_{x^{j}} \theta^{j}\left(x^{j}, \boldsymbol{v}(\boldsymbol{x})\right) \triangleq \nonumber\\ 	
	&col((\nabla_{x_{i}^{j}} \theta_{i}^{j}(x_{i}^{j}, z)+\frac{1}{n_{j}} \sum_{s=1}^{n_{j}} \nabla_{z^j} \theta_{s}^{j}(x_{s}^{j}, z))_{i \in \mathcal{S}_j})|_{z=\boldsymbol{v}(\boldsymbol{x}) },
\end{flalign}
\normalsize
where  the symbol $ \nabla_{z^j} $ means the gradient with respect to the aggregate quantity of cluster $ j $,
and the same as the symbol $ \nabla_{v^{j}} $ in what follows.
Similarly to \cite{gadjov2020single}, we define the pseudo-gradient mapping as
\begin{flalign} \label{pseduo-gradient}
	\Theta (\boldsymbol{x}) \triangleq
	\begin{pmatrix}
		\nabla_{x^1}\theta^1(x^1, \boldsymbol{v}(\boldsymbol{x}))  \\
		\vdots\\
		\nabla_{x^m}\theta^m(x^m,\boldsymbol{v}(\boldsymbol{x}))
	\end{pmatrix}.
\end{flalign}
\par For a multi-cluster aggregative game   (\ref{cluster-agg}),
$\boldsymbol{x}^*$ is a Nash equilibrium if   the following for each $ j\in \mathcal{V}$:
\begin{flalign}
	\theta^{j}\left(x^{j*}, \boldsymbol{v}(\boldsymbol{x}^*)\right)
	\leq  \theta^{j}\left(x^{j}, v^j(x^j),\boldsymbol{v}(\boldsymbol{x}^{-j*})\right).
\end{flalign}
\begin{lemma}(\cite{facchinei2007finite}, Proposition 1.4.2)
	If  $ \boldsymbol{x}^* $  is a Nash equilibrium  of (\ref{cluster-agg}), then it is a solution to the variational
	inequality problem VI($\Omega$,$ \Theta $),  i.e.,
	$ (\boldsymbol{x}-\boldsymbol{x}^*)^T\Theta(\boldsymbol{x}^*)  \geq 0 , \forall \boldsymbol{x}\in \Omega $,
	which is equivalent to
	\begin{flalign} \label{optimal-conditon}
		\boldsymbol{x}^*= \prod_{ \Omega} (\boldsymbol{x}^*-\gamma (\Theta  (\boldsymbol{x}^*,\mathbf{1}_n \otimes \boldsymbol{v}(\boldsymbol{x}^*)) )). 
	\end{flalign}
\end{lemma}
If $  \Omega = \mathbb{R}^{nq} $, it implies that $ \Theta(\boldsymbol{x}^*,\mathbf{1}_n \otimes  \boldsymbol{v}(\boldsymbol{x}^*))=\boldsymbol{0} $.
\par In order to facilitate the subsequent analysis, we denote by $ \hat{z}^j_{i}  \in \mathbb{R}^{mq} $, $ \forall i \in \mathcal{S}_j $,
 $ \hat{\mathbf{z}}_j=
(\hat{z}^j_{1},\dots,\hat{z}^j_{n_j})\in \mathbb{R}^{n_jmq}$.
Furthermore, the mappings  
$\Theta^j_{x^j} (x^j, \hat{\mathbf{z}}_j): \mathbb{R}^{n_jq} \times \mathbb{R}^{n_jmq} \rightarrow \mathbb{R}^{n_jq} $
and
$\Theta^j_{z^j} (x^j, \hat{\mathbf{z}}_j):
\mathbb{R}^{n_jq} \times \mathbb{R}^{n_jmq} \rightarrow \mathbb{R}^{n_j q} $
are defined
as follows
\small
\begin{flalign} 
	\Theta^j_{x^j} (x^j, \hat{\mathbf{z}}_j) \triangleq
	\begin{pmatrix}
		\nabla_{x^j_1}\theta^j_1(x^j_1 , \hat{z}^j_{1})  \\
		\vdots\\
		\nabla_{x^j_{n_j}}\theta^j_{n_j}(x^j_{n_j},\hat{z}^j_{n_j})
	\end{pmatrix}, \notag
\end{flalign}
\begin{flalign}
	\Theta^j_{z^j} (x^j,\hat{\mathbf{z}}_j) \triangleq
	\begin{pmatrix}
		\nabla_{z^j}\theta^j_1(x^j_1, \hat{z}^j_{1} )  \\
		\vdots\\
		\nabla_{z^j}\theta^j_{n_j}(x^j_{n_j},\hat{z}^j_{n_j})
	\end{pmatrix}. \notag
\end{flalign}
\normalsize
Note that  $ \Theta(\boldsymbol{x}):\mathbb{R}^{nq} \rightarrow  \mathbb{R}^{nq} $
and $  \Theta^j_{z^j} (x^j,\hat{\mathbf{z}}_j):\mathbb{R}^{n_jq} \times \mathbb{R}^{mq}    \rightarrow  \mathbb{R}^{mq}  $
are different mappings. 
Likewise, denote by $ \hat{\mathbf{z}} \triangleq (\hat{\mathbf{z}}_1,\dots, \hat{\mathbf{z}}_j,\dots, \hat{\mathbf{z}}_m) \in \mathbb{R}^{nm q}$, the mappings  $ \Theta_{x} (\boldsymbol{x}, \hat{\mathbf{z}})  $  and
$ \bar{\Theta}_z(\boldsymbol{x}, \hat{\mathbf{z}}) $ are defined 
as follows
\small
\begin{flalign} \label{notationTheta2}
	\Theta_{x} (\boldsymbol{x}, \hat{\mathbf{z}}) &\triangleq
	\begin{pmatrix}
		\Theta^1_{x^1} (x^1,\hat{\mathbf{z}}_1)\\
		\vdots\\
		\Theta^m_{x^m} (x^m,\hat{\mathbf{z}}_m)
	\end{pmatrix},  \nonumber	
	\Theta_{z} (\boldsymbol{x}, \hat{\mathbf{z}}) \triangleq
	\begin{pmatrix}
		\Theta^1_{z^1} (x^1,\hat{\mathbf{z}}_1)\\
		\vdots\\
		\Theta^m_{z^m} (x^m,\hat{\mathbf{z}}_m)
	\end{pmatrix},  \nonumber	 \\
	\bar{\Theta}_z(\boldsymbol{x}, \hat{\mathbf{z}}) &\triangleq
	\begin{pmatrix}
		\mathbf{1}_{n_1}\otimes \frac{1}{n_{1}} \mathbf{1}_{n_{1}}^{T} \Theta^1_{z^1}(x^1,\hat{\mathbf{z}}_1) \\
		\vdots \\
		\mathbf{1}_{n_m}\otimes \frac{1}{n_{m}} \mathbf{1}_{n_{m}}^{T} \Theta^m_{z^m}(x^m,\hat{\mathbf{z}}_m)
	\end{pmatrix} \in \mathbb{R}^{nq}. 
\end{flalign}
\normalsize
Based on (\ref{optimalcondition}), it can be verified that
\begin{flalign} \label{derived}
	\Theta (\boldsymbol{x})= \Theta_{x} (\boldsymbol{x},\hat{\mathbf{z}}) + \bar{\Theta}_{z} (\boldsymbol{x},\hat{\mathbf{z}})|_{z=v,\, \hat{\mathbf{z}}=\mathbf{1}_n \otimes \boldsymbol{v}(\boldsymbol{x})}.
\end{flalign}
Note that  $\Theta(\boldsymbol{x}^*,\mathbf{1}_n \otimes\boldsymbol{v}(\boldsymbol{x}^*)) $
is equivalent to   $ \Theta(\boldsymbol{x}^*)$. 
Furthermore,  we use $ \boldsymbol{v} $ and $ \boldsymbol{v^*} $ instead of $ \boldsymbol{v}(\boldsymbol{x}) $ and $ \boldsymbol{v}(\boldsymbol{x}^*) $ in what follows for the  brevity of notation.
\par From the above analysis, 
     we have that
     each agent needs the aggregate quantity vector
     $ (v^1(x^1),\dots,v^m(x^m))  $ and the aggregate gradient $  \sum_{s=1}^{n_{j}} \nabla_{z^j} \theta_{s}^{j}(x_{s}^{j}, z))_{i \in    \mathcal{S}_j}|_{z=\boldsymbol{v}(\boldsymbol{x})} $ when calculating the gradient of its objective function, cf., (\ref{optimalcondition}). 
     However, the aggregate quantity  vector and  the gradient  cannot be directly obtained by each agent. 
     In order to obtain the information and calculate the Nash equilibrium distributively by each agent, we design the 
     hierarchical communication scheme and present it in the following subsection.
\subsection{Communication Topology}\label{subsectionofcommunicationtopology}
\par 
Herein, we design a hierarchical communication scheme to obtain the aggregate quantity vector and the aggregate gradient of each cluster.
For the sake of clarity of exposition, we introduce the hierarchical scheme of three kinds of interactions: coordinator-coordinator, agent-agent, and coordinator-agent, as shown in Fig.1.
\begin{figure}[htbp]
	\centering
	\includegraphics[width=2.5in]{ 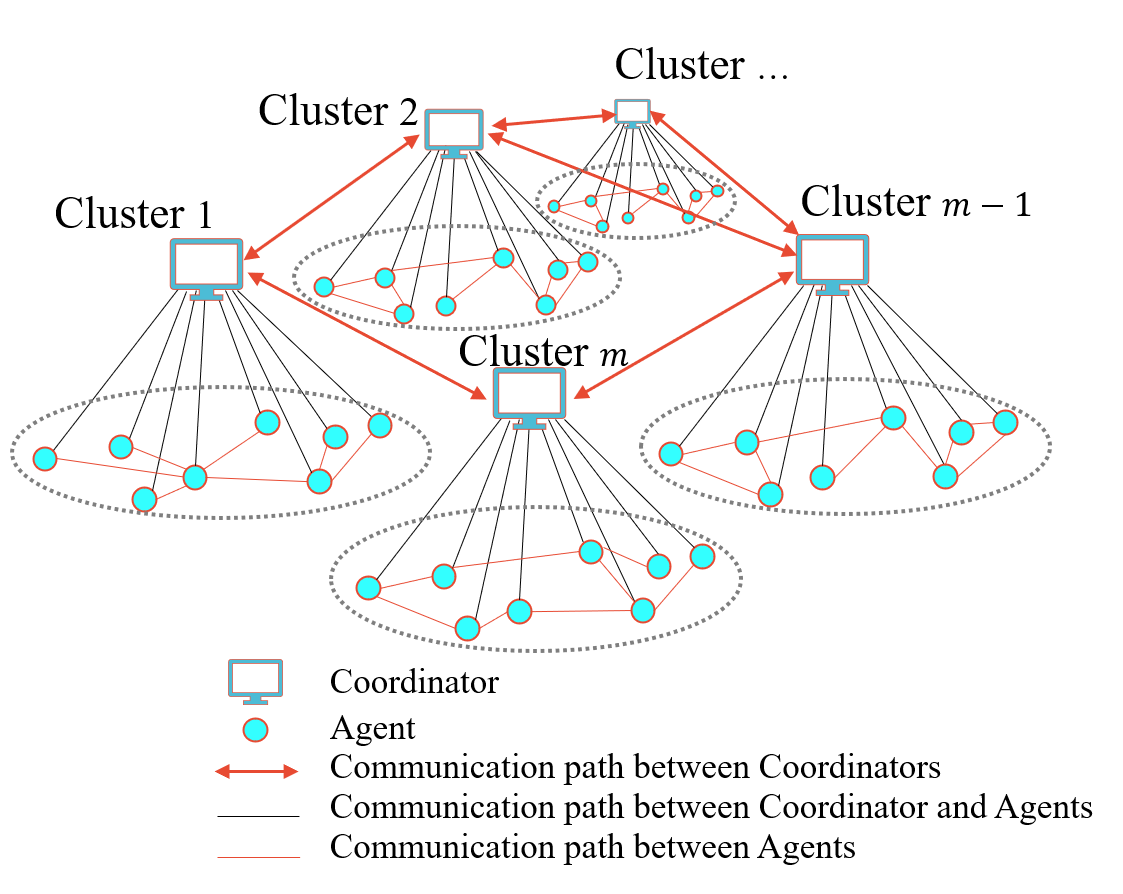}\\
	{Fig.1. Communication topology of the multi-cluster aggregative game. }
\end{figure}
\par The coordinator-coordinator interactions (the upper level)
are  guaranteed by an undirected graph.
Each coordinator owns an estimator to approximate $ (v^1(x^1),\dots,v^m(x^m))  $, e.g.,
for cluster $ j $, the estimate is denoted by
\[
\hat{\boldsymbol{v}}^j \triangleq (\hat{v}^{j(1)},\dots,\hat{v}^{j(h)},\dots,\hat{v}^{j(m)}),
\]
where $ \hat{v}^{j(h)} $ is coordinator $ j $'s estimate of  cluster $ h $'s aggregate quantity. 

\par The agent-agent interactions (the lower level) arise within each cluster, where each agent obtain the cluster's aggregate gradient by gradient tracking.  
For instance, in cluster $ j $, each agent $ i $  possesses a tracker $ \boldsymbol{y}^j_{i} \in \mathbb{R}^{n_jq}  $
to   dynamically track
$ \frac{1}{n_j}\sum_{i=1}^{n_j} \nabla_{v^j} \theta^j_i (x^j_{i}, \hat{\boldsymbol{v}}^j)   $ that is the global information of the whole cluster. 
\par Thus, the coordinator-agent interactions contain two processes: gathering and broadcasting.
The gathering process can be carried out through an interference graph (by observations) as described in \cite{yi2019operator}
to make sure the actual decisions of agents can be obtained by the coordinator.
The broadcasting process ensures the estimator value of each coordinator is received by its corresponding agents. 
\par Note that each agent only communicates to its neighbor agents within the cluster, and to its own coordinator which can be seen as a neighbor between clusters. 
\par For $ j\in \mathcal{V} $, we denote  by $ \mathcal{G}^j\triangleq(\mathcal{S}_j,\mathcal{E}_j)$  an undirected graph of cluster $ j $, where $ \mathcal{S}_j $ is the set of agents, $\mathcal{E}_j\subset \mathcal{S}_j\times \mathcal{S}_j$ is the edge set,
and $ W^j\triangleq[w^j_{is}]\in \mathbb{R}^{n_j\times n_j} $ is the adjacency matrix.
If $ w^j_{is} = w^j_{si} >0 $, it means agents $i,s$ within cluster $j$ can mutually exchange information,
and if  $ w^j_{is }=0 $, it means that agents $ i,s $ are not able to exchange information directly , cf., \cite{mesbahi2010graph}.
The set of neighbors of agent $ i $ in cluster $ j $ is defined as $ \mathcal{N}_{ji} = \{s| (i, s) \in  \mathcal{E}_{j}\} $.
We denote  $ d^j_{i}\triangleq\sum_{s=1}^{n_j}w^j_{is} $,
and $ Deg^j \triangleq diag((d^j_{i})_{i\in \mathcal{S}_j})\in\mathbb{R}^{n_j \times n_j} $.
The weighted Laplacian of $ \mathcal{G}^j $ is defined as $ L^j \triangleq Deg^j-W^j $.
For $ i,s\in \mathcal{S}_j $, if there exists a sequence of distinct nodes  $ i,i_1,\dots,i_p,s $ such that
$ (i,i_1) \in \mathcal{S}_j\times \mathcal{S}_j $,
$ (i_1,i_2) \in \mathcal{S}_j\times \mathcal{S}_j $,
$ \dots $,
$ (i_p,s) \in \mathcal{S}_j\times \mathcal{S}_j$,
then we call $ (i,i_1, \dots, i_p,s)  $ the undirected path between $ i $ and $ s $.
If there exists an undirected path between any $ i,s\in \mathcal{S}_j $, then $  \mathcal{G}^j $ is connected, cf., \cite{lei2016primal}.
In addition,  $ \mathcal{G}^0 $ is denoted as the coordinators' communication graph. 
The weighted matrix is
$ W^0 \triangleq[w^{jl}]\in \mathbb{R}^{m\times m}$, and Laplacian matrix is $ L^0 $.
The set of neighbors of coordinator $ j $ is $ \mathcal{N}^j = \{ l | (j,l) \in \mathcal{E}    \} $, $ j, l \in \mathcal{V} $,  where $ \mathcal{E}  $ is the edge set of $ \mathcal{G}^0 $.
We impose the following assumptions on the communication network.
\begin{assumption} \label{assumption_graph}
	Graphs $ \mathcal{G}^0 $, $ \mathcal{G}^1 $, $ \mathcal{G}^2 $, \dots, $ \mathcal{G}^m $,
	are undirected and connected.
	The adjacency matrix $ W^0 $ of $ \mathcal{G}^0 $ are double-stochastic 
	with positive diagonal elements, i.e.,
		$  W^0\mathbf{1}_m=\mathbf{1}_m $,
		$ \mathbf{1}^T_m W^0=\mathbf{1}_m^T $,
		$ w^{jj} >0 $, $ j\in \mathcal{V} $,
	while all adjacency matrices of clusters $ W^1 $, $ W^2 $, \dots, $ W^m $,
	are  double-stochastic, i.e.,	
	$ W^j \mathbf{1}_{n_j} =\mathbf{1}_{n_j} $,
	$ \mathbf{1}^T_{n_j} W^j=\mathbf{1}_{n_j}^T $,
	$ j\in \mathcal{V} $.
\end{assumption}
\par Assumption \ref{assumption_graph} is a typical setting for networked systems, cf., 
\cite{koshal2016distributed, nedic2017achieving}.
It can be fulfilled by using the Metropolis-Hastings rule, such as for cluster $ j $,
\begin{equation}
	w^j_{is}= 
	\begin{cases}
		\frac{1}{\max \{ d^j_{i}, d^j_{s} \}+1} & \text { if } i\neq s, \text{and} (i, s) \in \mathcal{E}_j, \\
		1-\sum_{s \in \mathcal{N}_{ji}} \frac{1}{\max \{ d^j_{i}, d^j_{s} \}+1} & \text { if } i=s, \\
		0 & \text { otherwise. }
	\end{cases} \notag
\end{equation}
The adjacency matrix $ W^0 $ can be defined analogously.
\section{Algorithm  Design} \label{section3} 
In this section, we propose a novel algorithm for Nash equilibrium seeking of the multi-cluster aggregative game (\ref{cluster-agg}).
The algorithm is shown as follows:
\begin{algorithm} [H]
	\caption{ Nash Equilibrium Seeking Algorithm for Multi-cluster Aggregative Games}\label{algorithm1}
	\textbf{Initialize}:
	For all  $  j,h \in \mathcal{V}  $, $ i \in \mathcal{S}_j $:
	$ x^j_{i,0}\in \mathbb{R}^q $,
	$v^j_{0} =\frac{1}{n_j} \sum_{i=1}^{n_j} x^j_{i,0} $, $  \hat{v}^{j(h)}_0 \in \mathbb{R}^q $,
	$ \hat{\boldsymbol{v}}^j_{0}=(\hat{v}^{j(0)}_0,\dots, \hat{v}^{j(h)}_0, \dots, \hat{v}^{j(m)}_0) $,
	$ \boldsymbol{y}^j_{i,0} =  \nabla_{v^{j}}  \theta^j_{i}(x_{i,0}^{j}, \hat{\boldsymbol{v}}^j_{0}) $. \\
	\textbf{Iteration}:  \\
	For every coordinator $ j  \in \mathcal{V}$, step $ k \geq 0 $,
	\small
	\begin{flalign}
		& v^j_{k} \leftarrow \frac{1}{n_j}\sum_{i=1}^{n_j}x^j_{i,k}:\text{gathers} \, x^j_{i,k}, \, \text{and} \, \text{calculates} \, v^j_{k}, \\
		&\textbf{for} \, h \,=\, 1\,\, \text{to} \, m \, \textbf{do} \notag \\
		&\quad  \hat{v}^{j(h)}_{k+1}=(\sum\nolimits_{s\in \mathcal{N}^j}w^{js}\hat{v}^{s(h)}_{k}) + \xi^{jh} w^{jh} (v^h_{k} -\hat{v}^{j(h)}_{k}), \label{key}\\
		&\textbf{end}   \notag \\
		&\hat{\boldsymbol{v}}^j_{k+1}=(\hat{v}^{j(1)}_{k+1},\dots, \hat{v}^{j(h)}_{k+1}, \dots, \hat{v}^{j(m)}_{k+1}),
	\end{flalign}
	\normalsize
	For every agent  $ i \in \mathcal{S}_j $ in cluster $  j\in \mathcal{V}  $, step $  k \geq 0  $,
	\small
	\begin{flalign}
		&\quad \, \, \, x^j_{i,k+1}= x_{i,k}^{j}-\gamma (\nabla_{x^j_{i}} \theta^j_{i}(x^j_{i,k},
		\hat{\boldsymbol{v}}^j_{k}   )+\boldsymbol{y}^j_{i,k}),  \label{key2}\\
		&\quad \, \, \, \boldsymbol{y}_{i,k+1}^{j}=
		(\sum\nolimits_{s \in \mathcal{N}_{j i}} w^{j}_{il} \boldsymbol{y}_{l,k}^{j})+
		(\nabla_{v^j} \theta^j_{i}(x_{i,k+1}^{j}, \hat{\boldsymbol{v}}^j_{k+1}) \nonumber  \\
		&\qquad \qquad \qquad \qquad \qquad \qquad \qquad -\nabla_{v^j} \theta^j_i(x_{i,k}^{j}, \hat{\boldsymbol{v}}^j_{k})),  \label{key3}
	\end{flalign}
	\normalsize
\end{algorithm}
where
$ v^j_k $,
$ \hat{v}^{j(h)}_{k} $,
$ \hat{\boldsymbol{v}}^j_{k} $,
$  x^j_{i,k}$,
and
$  \boldsymbol{y}_{i,k}^{j} $
denote
$ v^j $,
$ \hat{v}^{j(h)} $,
$ \hat{\boldsymbol{v}}^j $,
$  x^j_{i}$,
and
$  \boldsymbol{y}_{i}^{j} $
at iteration $ k $.
Moreover, $ \xi^{jh}$ is a positive constant fulfilling $ 0<\xi^{jh} <\frac{w^{jj}}{w^{jh}}$ in matrix $ \tilde{W}^ h,\, \forall w^{jh} \neq 0,\,\forall  h\in \mathcal{V} $ that is given by
\begin{flalign}
	\tilde{W}^h \triangleq
	\begin{pmatrix}
		& w^{11}- \xi^{1h} w^{1h}  &\dots  &w^{1m} \\
		&\vdots  &\ddots &\vdots\\
		&w^{m1}  & \cdots &  w^{mm}- \xi^{mh} w^{mh}
	\end{pmatrix}. \nonumber
\end{flalign}
\par For every coordinator $ j\in \mathcal{V} $ at iteration $ k $, it gathers the information of its corresponding agents and calculates the   aggregate value $ v^j_k=\frac{1}{n_j}\sum_{i=1}^{n_j}x^j_{i,k} $.
Furthermore, every coordinator renews its estimate $ \hat{v}^j_{k+1} $ of other clusters' aggregate values by \eqref{key}
based on the information gathered from its neighbors, i.e., $  \hat{v}^s_k, \forall s\in \mathcal{N}^j $.
Specifically,  if the estimated cluster like $ h $, already is the neighbor of  $ j $, then the actual value $ v^h_k $ can be directed transmitted to $ j $. 
The parameterized residual $ \xi^{jh}\omega^{jh}(v^h_k-\hat{v}^{j(h)}_k) $ is taken into account. 
However, if $ h $ is not a neighbor, i.e., $ \omega^{jh}=0 $, the residual is $ 0 $. 
After the updating process of all aggregative value estimators, each coordinator broadcasts its own estimates to all corresponding agents in that cluster. 
Finally,  agent $ i $ updates the strategies $ x^j_{i,k+1} $ and the tracker value $ \boldsymbol{y}^j_{i,k+1} $ by \eqref{key2} and \eqref{key3}  
based on the  information $ \hat{\boldsymbol{v}}^j_{k} $ obtained from the coordinator, respectively.
\par 
\begin{remark} \label{topology}
	Herein, we discuss another possible communication topologies when there is no communication within the cluster. 
	Namely, all agents communicate only with the coordinator (no agent-agent interactions), while agents not only receive information from the coordinator (coordinator-agent interactions) but also communicate with its neighbor nodes (agent-agent interactions).
	For a topology like this, all agents voluntarily send gradients to the coordinator to obtain the aggregate value calculated by the coordinator. 
	Note that for such a scheme, the interactions need \textit{both-way interactions} between the coordinator and agents, while agents only need to receive the information from the coordinator. 
	Nevertheless, for this kind of topology, agents need no interactions with the coordinator but communication with their neighbors in the cluster to obtain the aggregate gradient.
	\par However, in real engineering applications, agents might  not directly  share gradient information or deliver it to the coordinator  due to privacy concern,
      the cost function or gradient information is regarded as the sensitive information known to user that has to be kept confidential, 	
      such as in  \cite{nozari2016differentially}.
	In this case, it is very important to set up the communication structure among the agents within the cluster, so that each agent can estimate the gradient information of the whole cluster only by exchanging estimates with its neighbors without taking risks to transfer the privacy information to the coordinator.
\end{remark}
\section{Main Results} \label{section4}
\par In this section, we show the main results of the multi-cluster aggregative game by subsection preliminary results and subsection convergence analysis.  
\subsection{Preliminary Results}
\par In this subsection, we introduce the preliminary results including the
properties of $ W^j $, $ \tilde{W}^h $, the extended matrix $ \tilde{\mathsf{W}}^h $,
and the compact form of Algorithm \ref{algorithm1}, in order to further support the convergence prove.
\par Next, we give a lemma of a local cluster's gradient tacking method, cf., \cite{pu2021distributed}.
\begin{lemma}
	Suppose Assumption \ref{assumption_graph} holds, we have
	\small
	\begin{flalign}
		& \bar{\boldsymbol{y}}^j_{k}=\frac{1}{n_j}\sum_{i=1}^{n_j} \boldsymbol{y}^j_{i,k}
		= \frac{1}{n_j}  \sum_{i=1}^{n_j} \nabla_{v^j} \theta^j_i (x^j_{i,k}, \hat{\boldsymbol{v}}^j_{k}).   \nonumber
	\end{flalign}
\normalsize
\end{lemma}
\par \textit{Proof}:
Note that
\small
\begin{flalign}
	& \boldsymbol{y}^j_{k+1}= (W^j\otimes \boldsymbol{I}_q) \boldsymbol{y}^j_k
	+ ( \Theta_{v^{j}}^j (x^j_{k+1}, \hat{\mathbf{v}}^j_{k+1})
	- \Theta_{v^{j}}^j (x^j_{k}, \hat{\mathbf{v}}^j_{k})), \label{clusterlevelofy}
\end{flalign}
\normalsize
We multiply $ \mathbf{1}_{n_j q}^T/{n_j} $ to both sides of the (\ref{clusterlevelofy}), since $ \mathbf{1}_{n_j q}^T (W^j\otimes \boldsymbol{I}_q) =\mathbf{1}_{n_j q}^T $ in light of Assumption \ref{assumption_graph}.
Then, it suffices to have
\small
\begin{flalign}
	& \bar{\boldsymbol{y}}^j_{k+1}=\bar{\boldsymbol{y}}^j_{k}  +
	\frac{1}{n_j} \mathbf{1}_{n_j q}^T ( \Theta_{v^{j}}^j (x^j_{k+1}, \hat{\mathbf{v}}^j_{k+1})
	- \Theta_{v^{j}}^j (x^j_{k}, \hat{\mathbf{v}}^j_{k})).  \nonumber
\end{flalign}
\normalsize
By recalling the definition of  $\Theta_{v^{j}}^j (x^j, \hat{\boldsymbol{v}}^j))  $, we have
\small
\[
\frac{1}{n_j} \mathbf{1}_{n_j q}^T  \Theta_{v^{j}}^j (x^j_{k}, \hat{\boldsymbol{v}}^j_{k}) = \frac{1}{n_j}  \sum_{i=1}^{n_j} \nabla_{v^j} \theta^j_i (x^j_{i,k}, \hat{\boldsymbol{v}}^j_{k}).
\]
\normalsize
Thus, it follows that
\small
\begin{flalign}
	&\bar{\boldsymbol{y}}^j_{k+1}
	-\frac{1}{n_j}  \sum_{i=1}^{n_j} \nabla_{v^j} \theta^j_i (x^j_{i,k+1}, \hat{\boldsymbol{v}}^j_{k+1}) \nonumber\\
	& \qquad  \qquad =  \bar{\boldsymbol{y}}^j_{k}
	-\frac{1}{n_j}  \sum_{i=1}^{n_j} \nabla_{v^j} \theta^j_i (x^j_{i,k}, \hat{\boldsymbol{v}}^j_{k})	\nonumber  \\
	&\qquad  \qquad =  \bar{\boldsymbol{y}}^j_{0}
	-\frac{1}{n_j}  \sum_{i=1}^{n_j} \nabla_{v^j} \theta^j_i (x^j_{i,0}, \hat{\boldsymbol{v}}^j_{0}).	\nonumber
\end{flalign}
\normalsize
This together  with the initial values $ \boldsymbol{y}^j_{i,0}= \nabla_{v^j} \theta^j_i (x^j_{i,0}, \hat{\boldsymbol{v}}^j_{0})  $
implies  that $ \bar{\boldsymbol{y}}^j_{k}
= \frac{1}{n_j}  \sum_{i=1}^{n_j} \nabla_{v^j} \theta^j_i (x^j_{i,k}, \hat{\boldsymbol{v}}^j_{k})$.
\hfill $ \blacksquare $
\par In the following, we introduce the properties of the matrix $ \tilde{W}^h $ by two lemmas.
\begin{lemma} \label{lemma_of_spectral}
	Suppose Assumption \ref{assumption_graph} holds. 
    For all $j\in \mathcal{V}  $, if  $ 0 < \xi^{jh} < \frac{w^{jj}}{w^{jh}} $ holds, 
	then the spectral radius of $ \tilde{W}^h $, denoted by $ \rho(\tilde{W}^h) $, is less than 1, for all $  h\in \mathcal{V} $, i.e.,
	$ \rho(\tilde{W}^h)<1 $. Furthermore, there exists a norm $ \| \cdot \|_w $ such that $ \| \tilde{W}^h  \|_w < 1 $.
\end{lemma}
\par \textit{Proof}:
Due to Lemma 3 of  \cite{pang2020distributed}, it follows that $ \rho(\tilde{W}^h) <1 $.
Next, we introduce the following lemma to facilitate the proof.
\begin{lemma}(\cite{horn2012matrix}, Th.5.6.10) Let $  \tilde{W}^h  \in \mathbb{R}^{n\times n} $ and $ \epsilon>0 $ be given.
	There is  a matrix norm $ \| \cdot \|_w $ such that $ \rho ( \tilde{W}^h )  \leq \|   \tilde{W}^h  \|_w \leq \rho( \tilde{W}^h ) + \epsilon $.
\end{lemma}
\par By invoking the above lemma, it is straightforward to see that  $ \epsilon  $ can be any value, so we can set
$ \epsilon  $  sufficient small such that $ \rho( \tilde{W}^h ) + \epsilon <1 $, then there always exist a matrix norm
$ \| \cdot \|_w $ such that $ \| \rho ( \tilde{W}^h  )\|_w  <  \rho( \tilde{W}^h ) + \epsilon <1$, which facilitates the proof. 	
\hfill $\blacksquare$
\par Since $  \tilde{W} $ is symmetric, we can choose $ 2 $-norm directly, i.e., 
$ 	\|  \tilde{W}^h \|_2  <1  $.
\par 
We give the following definitions to recast the distributed algorithm \ref{algorithm1} into a collective form from an intuitive perspective. 
Recall that $ \hat{v}^{j(h)} $  is the coordinator $ j $'s  estimate to the aggregate quantity of cluster $ h $,
we can compactly denote all clusters' estimates of cluster $ h $ by
\[
\hat{v}^{:(h)} \triangleq (\hat{v}^{1(h)},\dots,\hat{v}^{j(h)},\dots,\hat{v}^{m(h)}).
\]
Furthermore,  the piled up vector from $ 1 $ to $ m $ is defined by $  \hat{V}= col(\hat{v}^{:(j)})_{j\in \mathcal{V}})  $.
\par For further analysis, here we define a matrix $ R^j $ to extract the estimate of cluster $ j $ from $\hat{V} $, i.e.,
\small
\begin{flalign}
	&R^j \triangleq
	\begin{pmatrix}
		\left[\boldsymbol{E}_{1} \right]_{11} &
		\dots&
		\left[\boldsymbol{0}_{q \times mq}\right]_{m1}& \\
		\vdots& \ddots& \vdots&\\
		\left[\boldsymbol{0}_{q \times mq}\right]_{1m }&
		\dots&
		\left[\boldsymbol{E}_m \right]_{mm}&
	\end{pmatrix},
	\nonumber
\end{flalign}
\normalsize
where $ [\boldsymbol{E}_j] \triangleq [\boldsymbol{0}_{q \times (<jq)}, \boldsymbol{I}_q, \boldsymbol{0}_{q\times(>jq)}] $, and there are $ m^2 $ blocks in $ R^j $, and the subscript of a block represent its positions
i.e., when the block's subscript is $ ij $, it means the block is in block row $ i $, block column $ j $.
Each block row has a
$ \left[  \boldsymbol{0}_{q \times (<jq)},\boldsymbol{I}_q,\boldsymbol{0}_{q\times(>jq)} \right] $ in respective position,
i.e.,
for block row $ j $, the block is in position $ jj $.
and the rest block row are all
$ \left[\boldsymbol{0}_{q \times mq}\right] $.
Thus, $ R^j \hat{V} =\hat{\boldsymbol{v}}^j $.
\par  For the brevity of notation , we denote
$ \boldsymbol{x}_k  \triangleq col((x^j_{k})_{j\in\mathcal{V}}) $,
$ \boldsymbol{v}_{k} \triangleq col((v^j_k)_{j\in \mathcal{V}}) $,
$ V_k \triangleq col((\mathbf{1}_m \otimes v^h_{k})_{h\in \mathcal{V}}) $,
$ \hat{\boldsymbol{v}}_{k} \triangleq col((\hat{v}^j_k)_{j\in \mathcal{V}}) $,
$ \hat{V}_k \triangleq col(\hat{v}^{:(h)}_k)_{h\in \mathcal{V}})  $,
$ \hat{\mathbf{v}}^j_k \triangleq col((\hat{v}^j_i)_{i\in \mathcal{S}_j})$,
$ \hat{\mathbf{v}}_k \triangleq col((\mathbf{v}^j_k)_{j \in \mathcal{V}})$,
$ \boldsymbol{y}^j_k \triangleq col((\boldsymbol{y}^j_{i,k})_{i\in \mathcal{S}_j})$,
$ \boldsymbol{Y}_k \triangleq col((\boldsymbol{y}^j_{k})_{j\in \mathcal{V}}) $,
$ \overline{\boldsymbol{Y}}_k \triangleq col(\mathbf{1}_{n_j}\otimes(\bar{\boldsymbol{y}}^j_{k})_{j\in \mathcal{V}}) $,
$ \tilde{W} \triangleq diag((\tilde{W}^j)_{j\in \mathcal{V}}) $,
$ W \triangleq diag((W^j)_{j\in \mathcal{V}}) $,
$ \boldsymbol{W} \triangleq W\otimes \boldsymbol{I}_q $,
and $ R \triangleq diag((R^j)_{j\in \mathcal{V}})  $.
\par In order to facilitate the calculation, we define a matrix $ M $ as follows
\small
\begin{flalign} \label{Mmatrix}
	M \triangleq
	\left(\begin{array}{ccc}
		\frac{1}{n_{1}} \mathbf{1}_{n_{1}}^{T} \otimes \boldsymbol{I}_q & \cdots & \boldsymbol{0} \\
		\vdots & \ddots & \vdots \\
		\boldsymbol{0} & \cdots & \frac{1}{n_{m}} \mathbf{1}_{n_{m}}^{T} \otimes \boldsymbol{I}_q
	\end{array}\right) \in \mathbb{R}^{mq \times nq},
\end{flalign}
\normalsize
and  we can get  $ n_M \triangleq \|  M  \| =\max_{j\in \mathcal{V}}{\sqrt{\frac{1}{n_j}}} $. Note that,  $ n_M $ is exploited to simplify the notation in what follows.
Moreover, we define
$ \boldsymbol{M} \triangleq   diag( ( \mathbf{1}_m  \otimes \frac{1}{n_{j}} \mathbf{1}_{n_{j}}^{T} \otimes \boldsymbol{I}_q  )_{j\in \mathcal{V}} )  \in  \mathbb{R} ^ {
	mmq \times nq}$.
By the definition of $ \boldsymbol{v} $, $ M $, $ V $, $ \boldsymbol{M}  $, it can be deduced that $\boldsymbol{v} = M \boldsymbol{x} $ and  $ V  =  \boldsymbol{M} \boldsymbol{x} $.
\par The compact form of Algorithm \ref{algorithm1}  is given as follows
\begin{flalign}
	         V_k =& \boldsymbol{M} \boldsymbol{x}_k,    \label{compactform_V}\\
			\hat{V}_{k+1}=&V_k-\left(\tilde{W} \otimes \boldsymbol{I}_{q}\right)\left(V_k-\hat{V}_{k}\right), \label{compactform_v}  \\
			\hat{\boldsymbol{v}}_{k+1}=&R(\mathbf{1}_m \otimes \hat{V}_{k+1}),  \label{compactform_Rx}\\
			\boldsymbol{x}_{k+1}=&\boldsymbol{x}_{k}
			-\gamma \left(\Theta_{x}\left(\boldsymbol{x}_{k}, \hat{\mathbf{v}}_{k}\right)+\boldsymbol{Y}_{k}\right),\label{compactform_x} \\
			\boldsymbol{Y}_{k+1}=&\boldsymbol{W} \boldsymbol{Y}_{\boldsymbol{k}}+\Theta_{v}\left(\boldsymbol{x}_{k+1},  \hat{\mathbf{v}}_{k+1}\right)-\Theta_{v}\left(\boldsymbol{x}_{k}, \hat{\mathbf{v}}_{k}\right). \label{compactform_y}
\end{flalign}
\par Suppose Assumptions \ref{assumption_func}, \ref{assumption_graph} hold. The Algorithm \ref{algorithm1} is equivalent to
	(\ref{compactform_V}), (\ref{compactform_v}), (\ref{compactform_Rx}), (\ref{compactform_x}), and (\ref{compactform_y}).
	Therein, the compact form of $ \hat{\boldsymbol{x}}_{k+1} $  and $ \overline{\boldsymbol{Y}}_{k+1} $  are straightforward
	by (\ref{key2}) and (\ref{key3}). Furthermore, for $ \hat{\boldsymbol{v}}_{k+1} $, in light of \ref{key}, we have
	\begin{flalign}
		\hat{v}_{k+1}^{:(h)}=\mathbf{1}_{m} \otimes v_{k}^{h}-(\tilde{W}^{h}\otimes \boldsymbol{I}_q)\left(\mathbf{1}_{m} \otimes v_{k}^{h}-\hat{v}_{k}^{:(h)}\right),  \label{estimatev:},
	\end{flalign}
	and we  stack vectors from  $ 1 $ to $ m $, which yields  (\ref{compactform_v}).
	For cluster $ j $ at time $ k $, every agent gets a copy estimate $ \hat{v}^j_k $ from coordinator $ j $ by receiving the broadcasting signal.
	Namely, for overall compatibility, the act of broadcasting is that every item $ \hat{v}^{j(h)}_k $  in vector $ \hat{v}^{:(h)}_k $ makes $ n_j  $  copies. As a result, it leads the dimensions of compact vector $ \hat{v}^{:(h)}_k $ extended from $ m $ to  $ n $. Denote the extended vector by $ \hat{\mathbf{v}}^{:(h)}_k $, then the extended (\ref{estimatev:}) is given by
	\begin{flalign}
		\hat{\mathbf{v}}^{:(h)}_{k+1}  =  \mathbf{1}_n \otimes v^h_k -  (\tilde{\mathsf{W}}^h  \otimes \boldsymbol{I}_q  )
		\left(  \mathbf{1}_n \otimes v^h_k  -  \hat{\mathbf{v}}^{:(h)}_{k}   \right),
	\end{flalign}
where  $ \hat{\mathbf{v}}^{:(h)}_{k} \triangleq col( (   \mathbf{1}_{n_j} \otimes \hat{v}^{j{(h)}}_{k}       )_{j  \in \mathcal{V}}      ) $,
and let $ \boldsymbol{e}^{n_j}=(1,0,\dots,0)\in \mathbb{R}^{n_j} $,
for item $ w^{js} $ in matrix $ \tilde{W}^h $ replaced by block $w^{js} \otimes \boldsymbol{1}_{n_j} \otimes {(\boldsymbol{e}^{n_s})}^T$
form matrix $ \tilde{\mathsf{W}}^h $, i.e.,
$ \tilde{\mathsf{W}}^h \triangleq [(w^{js} \otimes \boldsymbol{1}_{n_j} \otimes {(\boldsymbol{e}^{n_s}})^T)_{j,s \in \mathcal{V}}]  $.
Moreover,
by  elementary transformation of $ \tilde{\mathsf{W}}^h - \lambda \boldsymbol{I}_n $, we can get a matrix consisted of two nonzero blocks On the diagonal, while other elements of this matrix are all zeros.
One of the nonzero blocks is $ \tilde{W}^h - \lambda \boldsymbol{I}_m $, however, the other one is  $ -\lambda \boldsymbol{I}_{n-m} $.
According to the determinant properties of the partitioned matrix, we can get 
$\text{det}(\tilde{\mathsf{W}}^h - \lambda \boldsymbol{I}_n )  = \text{det} (\tilde{W}^h - \lambda \boldsymbol{I}_m)\text{det}( -\lambda \boldsymbol{I}_{n-m})$.
As a result, we can derive that the spectral radius of matrix $ \tilde{\mathsf{W}}^h $ is equivalent to $ \tilde{{W}}^h $,
i.e., $ \rho(\tilde{\mathsf{W}}^h) = \rho (\tilde{{W}}^h) $.
Additionally, invoking the properties of the Kronecker product, $ \rho (\tilde{W}^h \otimes \boldsymbol{I}_q) = \rho(\tilde{W}^h) $ is hold, and the same as $ \tilde{\mathsf{W}} $ which is given by
\begin{flalign}
	\rho(\tilde{\mathsf{W}}^h\otimes\boldsymbol{I}_q) = \rho (\tilde{\mathsf{W}}^h). \label{rhowotimes=rhow}
\end{flalign} 
\par To proceed with the analysis of the convergence, as the definitions of $ \hat{V}_k  $, $ V_k $ and $ \tilde{W}$,
we define the following variables:
$ \hat{\mathbf{V}}_k \triangleq col(\hat{\mathbf{v}}^{:(h)}_k)_{h\in \mathcal{V}})  $,
$ \mathbf{V}_k \triangleq col((\mathbf{1}_n \otimes \mathbf{v}^h_{k})_{h\in \mathcal{V}})  $,
and $ \tilde{\mathsf{W}} \triangleq diag((\tilde{\mathsf{W}}^j)_{j\in \mathcal{V}}) $.
For $ \hat{V}_k $, $ V_k $, $ \hat{\boldsymbol{v}}_{k} $, $\boldsymbol{v}_k $, $ \mathbf{v}_k $, $ \hat{\mathbf{V}}_k $, and $ \mathbf{V}_k $,
 we have
\( \| \hat{V}_k  -  V_k   \|
= \sqrt{\sum_{s=1}^m \sum_{j=1}^m \|  \hat{v}^{s(j)}_k -v^j_k \|^2} \),
\(
\|  \hat{\boldsymbol{v}}_{k} -\mathbf{1}_m \otimes \boldsymbol{v}_k       \|
=    \sqrt{\sum_{j=1}^m \sum_{s=1}^m \| \hat{v}^{j(s)}_k -v^s_k \|^2}
\).
Furthermore, it follows that
\(
\| \hat{V}_k  -  V_k   \|
=  \|  \hat{\boldsymbol{v}}_{k} -\mathbf{1}_m \otimes \boldsymbol{v}_k       \|
\), and
\begin{flalign} \label{vequalV}
\|  \hat{\mathbf{v}}_k   -   \mathbf{1}_n \otimes \boldsymbol{v}_k     \|
=&     \|   \hat{\mathbf{V}}_k  -  \mathbf{V}_k   \| \notag\\
=&\sqrt{ \sum_{j=1}^m \sum_{s=1}^m n_j \| \hat{v}^{j(s)}_k -v^s_k \|^2}. 
\end{flalign}
\subsection{Convergence Analysis}
\par  Next, we give assumptions on the mappings defined in (\ref{pseduo-gradient}) and (\ref{notationTheta2}) to facilitate the prove of the  convergence of Algorithm \ref{algorithm1}.
\begin{assumption} \label{assumption-pseduo}
	The following hold  for the  above defined   mappings associated with the problem (\ref{cluster-agg}):
	\begin{enumerate}
		\item
		The  mapping $ 	\Theta (\boldsymbol{x}) $  is strongly monotone  with $  \eta>0 $
		and Lipschitz continuous  with $ L>0 $ on $ \mathbb{R}^{nq} $, i.e.,
		\small
		\begin{flalign}
			&  \left(\Theta(\boldsymbol{x})-\Theta\left(\boldsymbol{x}'\right)\right)^{T}\left(\boldsymbol{x}-\boldsymbol{x}'\right)
			\geq \eta\left\|\boldsymbol{x}-\boldsymbol{x}'\right\|^{2},  \forall \boldsymbol{x}, \boldsymbol{x}' \in \mathbb{R}^{nq},  \nonumber \\
			& \|  \Theta(\boldsymbol{x})-\Theta(\boldsymbol{x}')  \|  \leq  L \|  \boldsymbol{x} - \boldsymbol{x}' \| ,  \forall \boldsymbol{x}, \boldsymbol{x}' \in \mathbb{R}^{nq}. \nonumber
		\end{flalign}
		\normalsize
		\item
		The mapping $ \Theta_{x}(\boldsymbol{x},\mathbf{z})+ \bar{\Theta}_{z}(\boldsymbol{x},\mathbf{z})$
		is Lipschitz continuous with constant $  L_{\Theta}>0  $  on $  \mathbb{R}^{nq }  \times \mathbb{R}^{nmq}  $, i.e.,
		\small
		\begin{flalign}
			&\left\|  \Theta_{x}(\boldsymbol{x},\mathbf{z})+ \bar{\Theta}_{z}(\boldsymbol{x},\mathbf{z})- 	
			\Theta_{x}(\boldsymbol{x}',\mathbf{z}')- \bar{\Theta}_{z}(\boldsymbol{x}',\mathbf{z}')  \right\|   \leq \nonumber \\
			&\quad   L_{\Theta} (\left\|  \boldsymbol{x}-\boldsymbol{x}^{\prime}   \right\|  +  \left\|  \mathbf{z}-\mathbf{z}^{\prime}   \right\| ),
			\forall\boldsymbol{x}, \boldsymbol{x}' \in \mathbb{R}^{nq},
			\forall   \mathbf{z}, \mathbf{z}'  \in  \mathbb{R}^{nmq}. \nonumber
		\end{flalign}
		\normalsize
		\item
		The mapping  $ \Theta_{z} (\boldsymbol{x},\mathbf{z}) $ is  Lipschitz continuous with constant $ L_v > 0 $  on  $ \mathbb{R}^{nq} \times \mathbb{R}^{nmq} $, i.e.,
		\small
		\begin{flalign}
			&\left\|  \Theta_{z} (\boldsymbol{x},\mathbf{z})-
			\Theta_{z} (\boldsymbol{x}^{\prime},\mathbf{z}^{\prime})  \right\|   \leq \nonumber \\
			&\quad  L_v (\left\|  \boldsymbol{x}-\boldsymbol{x}^{\prime}   \right\|  +  \left\|  \mathbf{z}-\mathbf{z}^{\prime}   \right\| ),
			\forall\boldsymbol{x}, \boldsymbol{x}' \in \mathbb{R}^{nq},
			\forall   \mathbf{z}, \mathbf{z}'  \in  \mathbb{R}^{nmq}. \nonumber
		\end{flalign}
		\normalsize
	\end{enumerate}
\end{assumption}
\par Assumption \ref{assumption-pseduo} is common in many literature on aggregative games and aggregative optimizations, such as \cite{fang2022distributed} and \cite{li2021distributed}. 
Furthermore, we give the following lemma that is also seen in  \cite{briceno2013monotone}.
\begin{lemma} \label{lemma_of_lipsmono}
    Suppose Assumption \ref*{assumption-pseduo} holds
	Then $ \|\boldsymbol{x}-\gamma \Theta (\boldsymbol{x}) - (\boldsymbol{x}'-\gamma \Theta(\boldsymbol{x}'))  \|
	\leq \sqrt{1-2\eta \gamma + L^2 \gamma^2} \|  \boldsymbol{x}  -  \boldsymbol{x}'    \|	$  hold, for all $ \boldsymbol{x},\boldsymbol{x}' \in \mathbb{R}^n $,
	where $ \gamma \in (0, \frac{2\eta}{L^2}) $.
\end{lemma}
\par \textit{Proof:}	Note that
\small
\begin{flalign}
	&	\|\boldsymbol{x}-\gamma \Theta (\boldsymbol{x}) - (\boldsymbol{x}'-\gamma \Theta(\boldsymbol{x}'))  \|^2  \nonumber \\
	&	\leq \| \boldsymbol{x} -  \boldsymbol{x}' \|^2
	-  2 \gamma \left\langle \boldsymbol{x} -  \boldsymbol{x}' , \Theta (\boldsymbol{x}) -  \Theta(\boldsymbol{x}')  \right\rangle
	+  \| \Theta (\boldsymbol{x}) -  \Theta(\boldsymbol{x}')   \|^2, \nonumber \\
	&   \leq^{(a)} (1-2\eta \gamma + L^2 \gamma^2) \|  \boldsymbol{x} -  \boldsymbol{x}'   \|^2, \nonumber
\end{flalign}
\normalsize
where $ (a) $  is invoking the strong monotonicity and the Lipschitz continuous properties of $ \Theta(\boldsymbol{x}) $.
\hfill $\blacksquare$
\par  To advance the proof,  we introduce a lemma in the following, and omit the proof since it is analogous to
Lemma 1 in \cite{xin2018linear} or Lemma 4  in \cite{li2021distributed}.
\begin{lemma} \label{lemma_rho}
	The following formulas hold under Assumption \ref{assumption_func},
	\begin{itemize}
		\item $ (W^j\otimes \boldsymbol{I}_q)  \boldsymbol{J}^j=\boldsymbol{J}^j (W^j\otimes\boldsymbol{I}_q)=\boldsymbol{J}^j $;
		\item $ \| (W^j\otimes \boldsymbol{I}_q) \boldsymbol{y}^j -\boldsymbol{J}^j \boldsymbol{y}^j  \| \leq \rho^j \| \boldsymbol{y}  - \boldsymbol{J}^j \boldsymbol{y}    \| $
		for any $ \boldsymbol{y}^j \in \mathbb{R}^{n_j q} $,
		where $ \rho^j \triangleq \|  (W^j\otimes \boldsymbol{I}_q)-\boldsymbol{J}^j    \| <1 $.\hfill $\blacksquare$
	\end{itemize}
\end{lemma}
\par We then present the following proposition, which is essential to demonstrate the convergence of Algorithm \ref{algorithm1}. 
\begin{proposition} \label{proposition 1}
	Suppose Assumptions  \ref{assumption_func}, \ref{assumption_graph}, and \ref{assumption-pseduo} hold.
	Furthermore,  if  $ \boldsymbol{x}^* $  fulfills the optimality condition (\ref{optimal-conditon}),
	then we have the following linear inequality system (LIS):
	\small
	\begin{flalign} \label{LIS1}
		\left(\begin{array}{c}
			\left\|\boldsymbol{x}_{k+1}-\boldsymbol{x}^{*}\right\| \\
			\left\|\hat{\mathbf{v}}_{k+1}-\mathbf{1}_{n} \otimes \boldsymbol{v}_{k+1}\right \| \\
			\|\boldsymbol{Y}_{k+1}-\overline{\boldsymbol{Y}}_{k+1}\|
		\end{array}\right) \leq G\left( \gamma \right)\left(\begin{array}{c}
			\left\|\boldsymbol{x}_{k}-\boldsymbol{x}^{*}\right\| \\
			\left\|\hat{\mathbf{v}}_{k}-\mathbf{1}_{n} \otimes \boldsymbol{v}_{k}\right \| \\
			\|\boldsymbol{Y}_{k}-\overline{\boldsymbol{Y}}_{k}\|
		\end{array}\right), 
	\end{flalign}
	\normalsize
	where the matrix $ G(\gamma) $ is defined as follows
	\small
	\begin{flalign} \label{matrix}
			G\left(\gamma\right) \triangleq 
		&	\begin{pmatrix}
			\Delta  &  L_{\Theta}  \gamma & \gamma \\
			G_1 \gamma & \sigma_w+  \tau  L_{\Theta}  \gamma & \tau  \gamma\\
			G_2 \gamma & L_{v} \sigma_s+L_{v} L_{\Theta}  \gamma
			& \rho+L_{v} \gamma
		\end{pmatrix},
	\end{flalign}
	\normalsize 
	with parameters defined by 
     $ \tau \triangleq \sqrt{n}n_M $,
	$ G_1 \triangleq \tau   L_{\Theta}(1+\tau  )  $,
	$ G_2 \triangleq  L_v L_{\Theta} + \tau L_v L_{\Theta}     $,
	$\Delta \triangleq \sqrt{1-2 \eta \gamma +  L^2 \gamma^2}$, for all $  \gamma \in (0, \frac{2\eta}{{L^2}}) $,
	$ \sigma_w \triangleq \|\tilde{\mathsf{W}}\otimes \boldsymbol{I}_q \| $,
	$ \sigma_s \triangleq \|\boldsymbol{I}_{n q}-\tilde{\mathsf{W}} \otimes \boldsymbol{I}_{q}\| $,
	and $ \rho \triangleq \| \boldsymbol{W} - \boldsymbol{J}  \| $.
\end{proposition}
\par {\textit{Proof:}} 	By (\ref{compactform_x}) and (\ref{optimal-conditon}),
	we have
	\small
	\begin{flalign} \label{TheThreeX}
				&\left\|\boldsymbol{x}_{k+1}-\boldsymbol{x}^{*}\right\| \nonumber \\
				& \leq\left\|\boldsymbol{x}_{k}-\boldsymbol{x}^{*}-\gamma\left(\left(\Theta_{x}\left(\boldsymbol{x}_{k}, \hat{\mathbf{v}}_{k}\right)+\boldsymbol{Y}_{k}\right)-\left(\Theta\left(\boldsymbol{x}^{*}, \mathbf{1}_{n} \otimes \boldsymbol{v}^{*}\right)\right)\right)\right\|  \nonumber \\
				& \leq \| \boldsymbol{x}_{k}-\boldsymbol{x}^{*}-\gamma(\Theta_{x}\left( \boldsymbol{x}_{k}, \mathbf{1}_n \otimes \boldsymbol{v}_{k} \right)+\bar{\Theta}_{v}\left(\boldsymbol{x}_{k}, \mathbf{1}_{n} \otimes \boldsymbol{v}_{k}\right)  \nonumber \\
				&\qquad -\Theta\left(\boldsymbol{x}^{*}, \mathbf{1}_{n} \otimes \boldsymbol{v}^{*}\right)+\Theta_{x}\left(\boldsymbol{x}_{k}, \hat{\mathbf{v}}_{k}\right)+\overline{\boldsymbol{Y}}_{k} \nonumber \\
				&\qquad \, -\Theta_{x}\left(\boldsymbol{x}_{k}, \mathbf{1}_{n} \otimes \boldsymbol{v}_{k}\right)-\bar{\Theta}_{v}\left(\boldsymbol{x}_{k}, \mathbf{1}_{n} \otimes \boldsymbol{v}_{k}\right)+\boldsymbol{Y}_{k}-\overline{\boldsymbol{Y}}_{k}) \|  \nonumber \\
				& \leq
				\|\boldsymbol{x}_{k}-\boldsymbol{x}^{*}   -  \gamma (\Theta_{x}(\boldsymbol{x}_{k}, \mathbf{1}_{n} \otimes \boldsymbol{v}_{k})
				+\bar{\Theta}_{v}\left(\boldsymbol{x}_{k}, \mathbf{1}_{n} \otimes \boldsymbol{v}_{k}\right) \notag \\
				&\qquad -\Theta\left(\boldsymbol{x}^{*}, \mathbf{1}_{n} \otimes \boldsymbol{v}^{*}\right)) \|
				+\gamma \| \Theta_{x}(\boldsymbol{x}_k,\hat{\mathbf{v}}_k) +\overline{\boldsymbol{Y}}_k  \notag \\
				&\qquad -\Theta_{x}(\boldsymbol{x}_k,\mathbf{1}_n \otimes\boldsymbol{v}_k)
				-\bar{\Theta}_v(\boldsymbol{x}_k,\mathbf{1}_n \otimes\boldsymbol{v}_k) \|
				+ \gamma \| \boldsymbol{Y}_k  -\overline{\boldsymbol{Y}}_k  \|\nonumber\\
				& \leq^{(a)} \Delta \left\|\boldsymbol{x}_{k}-\boldsymbol{x}^{*}\right\|+
				\gamma L_{\Theta} \left\|\hat{\mathbf{v}}_{k}-\mathbf{1}_{n} \otimes \boldsymbol{v}_{k}\right \| +\gamma\left\|\boldsymbol{Y}_{k}-\overline{\boldsymbol{Y}}_{k}\right\|,
	\end{flalign}
\normalsize
where $ \Delta \triangleq \sqrt{1-2\eta \gamma + L^2 \gamma^2} $. The first term of $ (a) $ follows from Lemma \ref{lemma_of_lipsmono}
, while the second term follows from Assumption \ref{assumption-pseduo}.
\par Next, for the second line of the LIS,  by  (\ref{vequalV}), we have
\small
\begin{flalign}  \label{V}
	&\| \hat{\mathbf{v}}_{k+1} - \mathbf{1}_n \otimes \boldsymbol{v}_{k+1}      \|
	=  \| \hat{\mathbf{V}}_{k+1}  -  \mathbf{V}_{k+1} \|  	\nonumber\\
	&\qquad \qquad \qquad =  \left\|   \mathbf{V}_k-(\tilde{\mathsf{W}}  \otimes \boldsymbol{I}_{q})
	(\mathbf{V}_k-\hat{\mathbf{\mathbf{V}}}_{k})   -  \mathbf{V}_{k+1}   \right \| \nonumber\\
	&\qquad \qquad \qquad \leq    \sigma_{w}   \|     \hat{\mathbf{v}}_{k} -\mathbf{1}_n \otimes \boldsymbol{v}_k     \|
	+    \left\|  \mathbf{V}_{k+1} -\mathbf{V}_k      \right \|,
\end{flalign}
\normalsize
where $  \sigma_w \triangleq \| \tilde{\mathsf{W}} \otimes  \boldsymbol{I}_q  \| $. By  (\ref{rhowotimes=rhow}), we can derive that $ \rho(\tilde{\mathsf{W}} \otimes  \boldsymbol{I}_q) <1 $ holds. 
Namely, $ \sigma_w \triangleq \| \tilde{\mathsf{W}} \otimes  \boldsymbol{I}_q  \| <1 $  holds.  
Moreover, for the second term of (\ref{V}), we can get
\small
\begin{flalign} \label{normofvsub}
	&\left\|  \mathbf{V}_{k+1} -  \mathbf{V}_k \right\| \notag \\
	&=\left\|\mathbf{1}_{n} \otimes \boldsymbol{v}_{k+1}-\mathbf{1}_{n} \otimes \boldsymbol{v}_{k}\right \|
	=\left\|\mathbf{1}_{n} \otimes\left(\boldsymbol{v}_{k+1}-\boldsymbol{v}_{k}\right)\right \| \nonumber \\
	& =\left\|\mathbf{1}_{n} \otimes\left(M \boldsymbol{x}_{k+1}-M \boldsymbol{x}_{k}\right)\right \|
	=\left\|\mathbf{1}_{n} \otimes M\left(\boldsymbol{x}_{k+1}-\boldsymbol{x}_{k}\right)\right \| \nonumber \\ &\leq\left\|\mathbf{1}_{n}\right\|\left\|M\left(\boldsymbol{x}_{k+1}-\boldsymbol{x}_{k}\right)\right \|
	\leq  \tau    \left\|\boldsymbol{x}_{k+1}-\boldsymbol{x}_{k}\right\|.
\end{flalign}
\normalsize
Thus, we have
\small
\begin{flalign} \label{TheThreeV}
	&\| \hat{\mathbf{v}}_{k+1}-\mathbf{1}_{n} \otimes  \boldsymbol{v}_{k+1}  \|   \notag \\    
	& \leq \tau   L_{\Theta} \gamma(1+\tau  )
	\left\|\boldsymbol{x}_{k}-\boldsymbol{x}^{*}\right\|    
	+\tau   \gamma\left\|\boldsymbol{Y}_{k}-\overline{\boldsymbol{Y}}_{k}\right\|   
	\nonumber \\  
	& +\left(\tau   \gamma L_{\Theta} +
	\sigma_w\right)\left\|\hat{\mathbf{v}}_{k}-\mathbf{1}_{n} \otimes \boldsymbol{v}_{k}\right \|. 
\end{flalign}
\normalsize
\par In the following, we aim to derive the third line of LIS.
More succinctly, denote
      $ \boldsymbol{W}^j\triangleq W^j\otimes \boldsymbol{I}_q $,
       $ J^{j} \triangleq \frac{1}{n_{j}} \mathbf{1}_{n_j}\mathbf{1}_{n_j}^{T}$,
       $ \boldsymbol{J}^j \triangleq J^j \otimes \boldsymbol{I}_q  $,
       $ \boldsymbol{J} \triangleq diag((\boldsymbol{J}^j)_{j\in \mathcal{V}}) $,
       $ \boldsymbol{I}_n \triangleq diag((\boldsymbol{I}_{n_j})_{j\in \mathcal{V}})$,
       and $\boldsymbol{ W} $ be as defined in (\ref{compactform_y}),
in light of  Lemma $ \ref{lemma_rho} $, it is easy to verify that 
$ \boldsymbol{W} \boldsymbol{Y}_{k}-\boldsymbol{J} \boldsymbol{Y}_{k}=(\boldsymbol{W}-\boldsymbol{J})\left(\boldsymbol{Y}_{\boldsymbol{k}}-\boldsymbol{J} \boldsymbol{Y}_{k}\right) $
hold. 
Therefore,
\small
\begin{flalign}
		&\left\|\boldsymbol{Y}_{k+1}-\overline{\boldsymbol{Y}}_{k+1}\right\|  \notag\\
		&\leq\left\|\boldsymbol{W} \boldsymbol{Y}_{k}-\boldsymbol{J} \boldsymbol{Y}_{k}+(\boldsymbol{I}-\boldsymbol{J})
		\left(\Theta_{v}(\boldsymbol{x}_{k+1}, \hat{\mathbf{v}}_{k+1})-\Theta_{v}(\boldsymbol{x}_{k}, \hat{\mathbf{v}}_{k})\right)\right\| \nonumber \\
		& \leq\left\|(\boldsymbol{W}-\boldsymbol{J})\left(\boldsymbol{Y}_{k}-\boldsymbol{J} \boldsymbol{Y}_{k}\right) \right\| \notag \\
		&\qquad\quad+ \left\| (\boldsymbol{I}-\boldsymbol{J})\right\| \left\|( \Theta_{v}(\boldsymbol{x}_{k+1}, \hat{\mathbf{v}}_{k+1})-\Theta_{v}(\boldsymbol{x}_{k}, \hat{\mathbf{v}}_{k}))\right\|, \label{Y-Ybar}
\end{flalign}
\normalsize
by  $ \rho \triangleq \| \boldsymbol{W} - \boldsymbol{J}  \| = max\{ \rho^j \} <1 $,
 $ \| \boldsymbol{I}  -  \boldsymbol{J}  \| = 1 $,  and  Assumption \ref{assumption-pseduo}, we have
\small
\begin{flalign}  \label{normofY}
	&\left\|\boldsymbol{Y}_{k+1}-\overline{\boldsymbol{Y}}_{k+1}\right\| \leq \rho\left\|\boldsymbol{Y}_{k}-\overline{\boldsymbol{Y}}_{k}\right\|+L_{v}\left(\left\|\boldsymbol{x}_{k+1}-\boldsymbol{x}_{k}\right\|\right.
	\nonumber \\
	&\qquad  \qquad \qquad  \qquad +\left.\left\|\hat{\mathbf{v}}_{k+1}-\hat{\mathbf{v}}_{k}\right\|\right).
\end{flalign}
\normalsize 
In the following,  
we will separately analyze the second and third terms on the right-hand  of (\ref{normofY}). 
In light of  Assumption \ref{assumption-pseduo},
we have 
\small
\begin{flalign} \label{xkplus1minusxk}
	&\| \boldsymbol{x}_{k+1} -\boldsymbol{x}_k\|  \nonumber \\
	& \leq  \gamma \|  ( \Theta_{x} (\boldsymbol{x}_k, \hat{\mathbf{v}}_k ) )+\overline{\boldsymbol{Y}}_k
	- \Theta(\boldsymbol{x}^*,\mathbf{1}_n \otimes \boldsymbol{v}^*) + \boldsymbol{Y}_k -\overline{\boldsymbol{Y}}_k \|  \nonumber  \\
	&  \leq  L_{\Theta} \gamma\left\|\boldsymbol{x}_{k}-\boldsymbol{x}^{*}\right\|
	+  L_{\Theta}    \gamma  \left\|\hat{\mathbf{v}}_{k}-\mathbf{1}_{n} \otimes \boldsymbol{v}^*\right \|
	+\gamma\left\|\boldsymbol{Y}_{k}-\overline{\boldsymbol{Y}}_{k}\right\|  \nonumber \\
	& \leq  L_{\Theta} \gamma\left\|\boldsymbol{x}_{k}-\boldsymbol{x}^{*}\right\|
	+  L_{\Theta}    \gamma  \left\|\hat{\mathbf{v}}_{k} -\mathbf{1}_n \otimes \boldsymbol{v}_k \right \|
	+  \gamma\left\|\boldsymbol{Y}_{k}-\overline{\boldsymbol{Y}}_{k}\right\| \nonumber \\
	&+ L_{\Theta}    \gamma \|\mathbf{1}_n \otimes \boldsymbol{v}_k -\mathbf{1}_{n} \otimes \boldsymbol{v}^* \|,
\end{flalign}
\normalsize
and by the definition of $ M $ in (\ref{Mmatrix}), i.e., $  \boldsymbol{v} = M \boldsymbol{x} $, it yields
$  \|\mathbf{1}_n \otimes \boldsymbol{v}_k -\mathbf{1}_{n} \otimes \boldsymbol{v}^* \|
\leq \| \mathbf{1}_n \otimes ( M\boldsymbol{x}_k - M\boldsymbol{x}^* )    \| \leq  \tau  \|  \boldsymbol{x}  -   \boldsymbol{x}^*    \| $.
Thus, (\ref{xkplus1minusxk}) is equivalent to
\small
\begin{flalign}
&\| \boldsymbol{x}_{k+1} -\boldsymbol{x}_k\|  \nonumber \\
&\leq   (L_{\Theta} \gamma + \tau L_{\Theta}    \gamma)   \left\|\boldsymbol{x}_{k}-\boldsymbol{x}^{*}\right\|
+  \gamma\left\|\boldsymbol{Y}_{k}-\overline{\boldsymbol{Y}}_{k}\right\| \notag \\
& +  L_{\Theta}    \gamma  \left\|\hat{\mathbf{v}}_{k} -\mathbf{1}_n \otimes \boldsymbol{v}_k \right \| .\nonumber
\end{flalign}
\normalsize
Furthermore, it follows from  (\ref{rhowotimes=rhow}) and  (\ref{vequalV}) that
\small
\begin{flalign} \label{vhatkplus1-vhatk}
	 \|\hat{ \mathbf{v}}_{k+1} -\hat{ \mathbf{v}}_{k}  \|
	  &=  \|  \hat{\mathbf{V}}_{k+1}  -\hat{\mathbf{V}}_k   \|    \nonumber \\
	 & =\| \mathbf{V}_k-(\tilde{\mathsf{W}} \otimes \boldsymbol{I}_{q})(\mathbf{V}_k-\hat{\mathbf{V}}_{k}) - \hat{\mathbf{V}}_k  \| \nonumber \\
	 & \leq \| \boldsymbol{I}_{nq} - \tilde{\mathsf{W}} \otimes \boldsymbol{I}_{q}  \|   \| \mathbf{V}_k-\hat{\mathbf{V}}_{k}  \| \nonumber \\
	 & \leq \sigma_{s}     \| \hat{\mathbf{v}}_{k} -\mathbf{1}_n \otimes \boldsymbol{v}_k  \|.
\end{flalign}
\normalsize 
\par Above all, due to (\ref{normofY}), (\ref{xkplus1minusxk}), and (\ref{vhatkplus1-vhatk}), it can be verified that
\small
\begin{flalign}
	&\left\|\boldsymbol{Y}_{k+1}-\overline{\boldsymbol{Y}}_{k+1}\right\| \leq
	(\rho+L_v\gamma)\left\|\boldsymbol{Y}_{k}-\overline{\boldsymbol{Y}}_{k}\right\|  \nonumber \\
	&\qquad   + (L_{v} L_{\Theta}\gamma + \tau   L_v L_{\Theta}   \gamma)
	\| \boldsymbol{x}_{k}-\boldsymbol{x}^* \|  \nonumber \\
	&\qquad     + ( L_v L_{\Theta}   \gamma + L_v  \sigma_{s}    )
	\| \hat{\mathbf{v}}_{k} - \mathbf{1}_n \otimes \boldsymbol{v}_{k} \|.  \label{TheThreeY}
\end{flalign}
\normalsize
\par By combining  (\ref{TheThreeX}), (\ref{TheThreeV}), and (\ref{TheThreeY}), we conclude that the LIS proposed in Proposition \ref{proposition 1} holds, which further implies that 
 (\ref{matrix}) holds.
	\hfill $\blacksquare$
\begin{lemma}(\cite{horn2012matrix},Th.6.3.12) \label{leftrighteigen}
	Let $ G(\gamma)  $ be differential at $ \gamma=0 $. Assume that $ \lambda $ is an algebraically
	simple eigenvalue of $ G(0) $, and $ \lambda(\gamma) $ is an eigenvalue of $ G(\gamma) $,
	for small $ \gamma $,  $ \lambda(0) = \lambda $ holds.
	Let $ w $ and $ v $ be a left eigenvector and a right eigenvector of $ G $, respectively. 
	Thus, the following equality holds:
	\[
	\lambda'(0) = \frac{w^T G'(0) v  }{w^T v}.
	\] 
\end{lemma}
\begin{proposition} \label{proposition_2}
	There exists a positive step-size $ \gamma $ near $ 0 $,  which results in the spectral radius of $ G(\gamma) $ strictly less than 1.
	The bound of $ \gamma $ can be estimated by solving the $ \text{det} (\boldsymbol{I}_3- G(\gamma))=0 $. Namely,
	let $ \gamma_s $ be the smallest positive root of $ \text{det} (\boldsymbol{I}_3- G(\gamma))=0 $, then for every $   0 < \gamma < \text{min}\{\gamma_s, \frac{2\eta}{L^2}\} $, $ \rho(G(\gamma))<1 $ holds.
\end{proposition}
\par {\textit{Proof:}}
Firstly, we prove the existence of the $ \gamma $.
Let $ \gamma=0 $, we have
\begin{flalign}
	G(0)=\begin{pmatrix}
		1& 0& 0 \\
		0&  \sigma_{w}& 0 \\
		0& L_v   \sigma_{s}& \rho
	\end{pmatrix}.
\end{flalign}
We can observe that $ \rho (G(0)) =1 $, since $ 0< \sigma_{w} <1$ and $ 0< \rho <1$.
Accordingly, we have  $ w = [1, 0, 0]^T$ and $ v=[1,0,0]^T $ to be the left and right eigenvector of $ G  $ respectively.
Since the existence of $ \gamma $ depends on how the eigenvalue $ \lambda(\gamma) $ of $ G(\gamma) $ changes near $ 0 $,
we need to take the derivative of $ \lambda(\gamma) $.
Also, by Theorem 1 of \cite{greenbaum2020first},  $ \lambda(\gamma) $ is unique at the neighborhood of $ \gamma=0 $.
In light of Lemma \ref{leftrighteigen}, we have
\small
\begin{flalign}
	&\left.	\frac{d\lambda(\gamma)}{d \gamma} \right|_{\gamma=0}  =  	\frac{ w^T \left.\frac{dG(\gamma)}{d \gamma} \right|_{\gamma=0} v}{w^Tv}
	\left.=\frac{-2\eta + 2L \gamma }{2 \sqrt{1-2\eta \gamma + L^2 \gamma^2 }} \right|_{\gamma=0}=-\eta. \nonumber
\end{flalign}
\normalsize
Note that $ G(\gamma) $ is continuous with respect to $ \gamma $, and so does $ \lambda(\gamma) $.
Since  $ -\eta < 0 $,  $ \lambda(\gamma)  $ decreases while $ \gamma $ increases.
Recall that $ \rho(G(0))=1 $, then we can conclude that
when $ \gamma $  increases from $ 0 $  to a positive sufficient small value,  $ \rho <1 $.
Namely, there always exists $ \gamma >0  $ such that $ \rho(G(\gamma)) <1 $.
Secondly,
by Perron-Frobenius theorem, $ G(\gamma) $ is primitive and $ \rho(G(\gamma)) $  is a simple and biggest eigenvalue of $ G(\gamma) $.
Since the determinant is equal to the product of the eigenvalues, $ \text{det}(\boldsymbol{I}_3 - G(\gamma^*) ) =0$ implies $\rho(G(\gamma^*))=1$.
It can be deduced that for the smallest positive solution of this equation $ \gamma_s $,  $ \text{det}(\boldsymbol{I}_3 - G(\gamma_s) ) =0$ hold.
Furthermore, for every $   0 < \gamma < \text{min}\{\gamma_s, \frac{2\eta}{L^2}\} $, $ \rho(G(\gamma))<1 $ holds.
\hfill $ \blacksquare $
\begin{theorem} \label{theorem}
	Suppose Assumptions \ref{assumption_func}, \ref{assumption_graph}, and \ref{assumption-pseduo}  hold.
	When $  0 < \gamma < \text{min}\{\gamma_s, \frac{2\eta}{L^2}\}$  holds,  $ \boldsymbol{x}_k $ generated by Algorithm \ref{algorithm1}  converges to
	a Nash equilibrium $ \boldsymbol{x}^* =(x^{1*},\dots, x^{m*}) $ of problem (\ref{cluster-agg})
	at a linear convergence rate,
	and $ \boldsymbol{x}^* $ fulfills the optimality condition  (\ref{optimal-conditon}).
	Furthermore, all coordinators' estimates of aggregate quantities reach a consensus when $ k \rightarrow \infty $, i.e.,
	\begin{flalign} \label{nashequilibrium}
		\lim_{k \rightarrow \infty} \hat{v}^{j}_k=\lim_{k \rightarrow \infty} \hat{v}^{h}_k
		= \lim_{k \rightarrow \infty} \boldsymbol{v}_k =\boldsymbol{v}^* , \, \forall j,h\in \mathcal{V},
	\end{flalign}
	where $  \hat{v}^{j}_k \triangleq  \hat{v}(x^{j}_k )$,
	$  \hat{v}^{h}_k \triangleq  \hat{v}(x^{h}_k ) $,
	 $ \boldsymbol{v}_k  \triangleq \boldsymbol{v} ( \boldsymbol{x}_k )   $,
	 $ \boldsymbol{v}^* \triangleq  \boldsymbol{v}(\boldsymbol{x}^*) $.
\end{theorem}
\par {\textit{Proof:}}
In light of  Proposition  \ref{proposition 1} and Proposition \ref{proposition_2}, we have
\[
\lim_{k \rightarrow \infty} \hat{ \mathbf{v}} _k =  \lim_{k \rightarrow \infty} \mathbf{1}_n \otimes  \boldsymbol{v}_k
=  \mathbf{1}_n \otimes  \boldsymbol{v}^*,
\]
which represents that the copies of coordinators' aggregative mappings' estimates reach consensus, and further implies that
the coordinators' estimates reach consensus.
\par Due to the strong monotonicity in Assumption \ref{assumption_graph},
there exists a unique Nash equilibrium of (\ref{cluster-agg})
, and it satisfies the optimal condition (\ref{optimal-conditon}).
Furthermore, in conjunction of Proposition   \ref{proposition 1} and  \ref{proposition_2}, we have that $ \boldsymbol{x}_k \rightarrow \boldsymbol{x}^*  $  with a linear rate when $ k\rightarrow  \infty$.
%
\hfill  $ \blacksquare $
\begin{remark}
	From the above analysis, it can be deduced that the scenario discussed in Remark \ref{topology}, where there is no agent-agent communication within each cluster,  is a special case of the topology shown in Fig.1. Namely, Algorithm \ref{algorithm1} is still suitable for such a topology.
\end{remark}
\par Since $ \gamma_s $ in Proposition \ref{proposition_2} is with high calculation expense, we utilize the following propositions to 
obtain a slightly strict upper-bound $ \gamma'_s $.
\begin{proposition}  \label{corollary}
	Suppose Assumptions 1-3 hold. 
	We utilize $1-\Xi\gamma$ to replace $\Delta $ in (\ref{matrix}), 
	and it forms a matrix $M(\gamma)$, 
	where $ \Xi \triangleq  \frac{1- \sqrt{ 1+ \frac{(a^2-2a)\eta^2}{L^2}}}{\frac{a\eta}{L^2}}$, $ a\in (0, 2)  $.
    Then we have
	\small 
	\begin{flalign}
		\rho (G(\gamma) )  <  \rho ( M(\gamma)), \forall \gamma \in (0, 
		\frac{a\eta}{L^2}].
	\end{flalign} 
\normalsize
Furthermore, if $ \gamma'_s $  is the smallest positive solution of $ \text{det}(\boldsymbol{I}_3 - M(\gamma)) =0$,  
then $  \rho ( G(\gamma) ) <1$ holds, for all $  \gamma \in \text{min} \{ r'_s , \frac{a\eta}{L^2}   \}, a\in (0,2)$, where 
$ r'_s $  is given by 
\small
\begin{flalign}
	\gamma'_s   =  \frac{\Xi (1-\rho )(1- \sigma_{w})  }{(G_1+\tau\Xi)(L_{\Theta}-\rho L_{\Theta} + L_v \sigma_s) + 
		(1-\sigma_{w}) ( \Xi L_v + G_2)}, \notag
\end{flalign} 
\normalsize
where $ G_1 $, $ G_2 $, and $ \Xi $ are defined in Proposition \ref{proposition 1} and \ref{corollary}.
\end{proposition}
\par {\textit{Proof:}}
By the definition of $ \Xi $ and $ a $, we have that $ \Xi $  is a constant when  $ a \in (0,2) $ is given. 
Furthermore, $1 - \Xi\gamma$ is a function of $\gamma$, and it always has a greater function value than another function $\Delta(\gamma)$   when $ \gamma$  is in $ (0, \frac{a\eta}{L^2}] $.
Notably, $ G(\gamma) $ and $ M(\gamma) $ are nonnegative.
By Corollary 8.1.19 of \cite{horn2012matrix}, we have $  \rho ( G(\gamma))  <  \rho (M(\gamma)) $ since $ \Delta(\gamma) <  1-\Xi \gamma $,
which implies that if a proper $ \gamma $ fulfills $ \rho (M(\gamma)) <1 $, then it must have $ \rho (G(\gamma)) <1 $ holds.
Additionally, $ \gamma'_s $ can be calculated by solving $ \text{det}(\boldsymbol{I}_3 - M(\gamma)) =0 $. 
The proof is complete.
\hfill $ \blacksquare $
\section{Application to The Energy Internet} \label{section5}
\par In this section, we present  a numerical simulation of an Energy Internet system.
\par
In an Energy Internet system, an energy subnet corresponds to a cluster in the game, while an energy hub corresponds to an agent.
As shown in Fig.2, all energy hubs are physically linked together by a unified energy bus.
For each energy subnet, there exists a coordinator gathering  information from hubs.
Each coordinator is in charge of a certain number of energy hubs' information transmitting,
meanwhile, they can also communicate  mutually at the same time.
In addition, we hypothesize that each hub can communicate with adjacent hubs in the same subnet to obtain estimates of the aggregate functional values of the entire subnet, 
which are referred to as private information  in  literature (see e.g., \cite{zhu2017differentially}) while avoiding direct transmission of these values.
 \begin{figure}[htbp]
 	\centering
 	\includegraphics[width=3.5in]{ 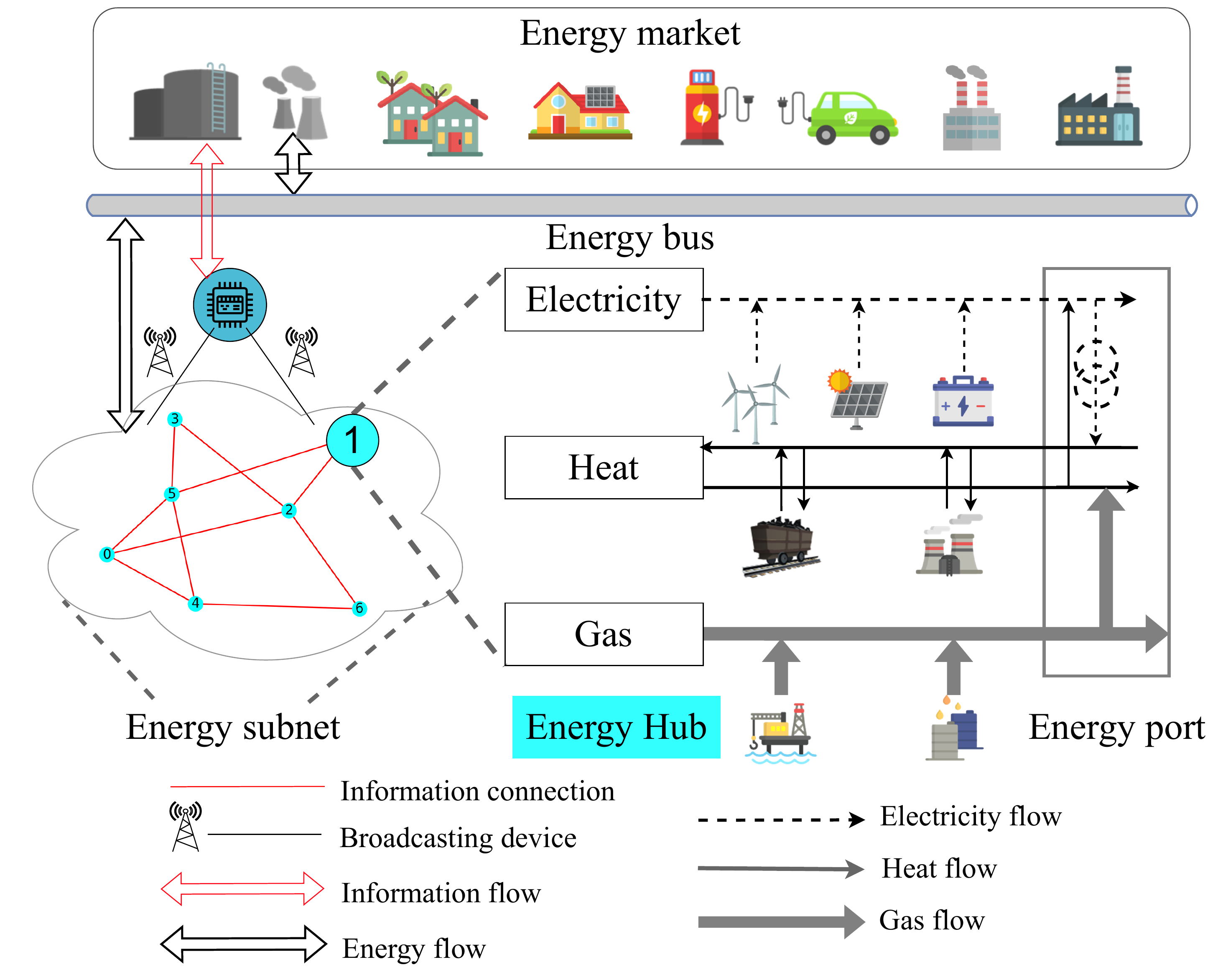}\\
 	{Fig.2. The schematic of an Energy Internet system and energy hubs. }	
 \end{figure}
\subsection{Case 1: Small-Scale Energy Subnets}
\par Herein, we consider a game with 4 energy subnets in different regions (communities) of 7, 8, 9, 10 hubs (unified interfaces) respectively.
For each $ j\in {1,2,3,4}, i\in \mathcal{S}_j $, without loss of generality,
set $\gamma=0.003 $, $  0< \xi^{jh} <\frac{w^{jj}}{w^{jh}} $, every element in diagonal matrix $ Q^j_i $  is randomly drawn from $ [2,4] $.
The same as $ b^j_i $, $ d^j_i $, and $ c^j_i  $, elements are randomly drawn from $ [-3, 1] $, $ [1,3] $, and $ [4,5] $, respectively. Moreover,  $ C^j_i \in \mathbb{R}^{3\times3} $ is a diagonal matrix, and the elements on the diagonal take different values corresponding to different energy policies, where the element values are randomly drawn from $ (0, 1] $.

\par The trajectories of relative error
 $ \frac{\| \boldsymbol{x}_k - \boldsymbol{x}^* \|}{\|  \boldsymbol{x}_0 - \boldsymbol{x}^*     \|} $ 
 of Algorithm 1 and Algorithm in \cite{zimmermann2021solving} are  depicted in Fig.4.
Recall that the Algorithm in \cite{zimmermann2021solving} require an extra communication graph that is independent of all other inner-cluster communication subgraphs. 
We can see that although both $ \boldsymbol{x}_k  $ converge to Nash equilibrium at a linear convergence rate,
Algorithm 1 can reach a better rate compared to \cite{zimmermann2021solving}.  
\par In addition, denote the  $ \hat{v}^j[1] $ the first element of $ \hat{v}^j $,
i.e., the estimate of aggregate electricity flow of cluster $ j $.
As shown in  Fig.4,  we present the trajectories of electricity strategy $ \hat{\boldsymbol{v}}[\mathbf{1}] $, from which it is seen that  all coordinators' estimates will eventually reach consensus, i.e., $ \hat{v}^{j}=\hat{v}^{h},\forall j,h \in \mathcal{V}  $.
\begin{figure}[htbp]
	\centering
	\includegraphics[width=3.5in]{ 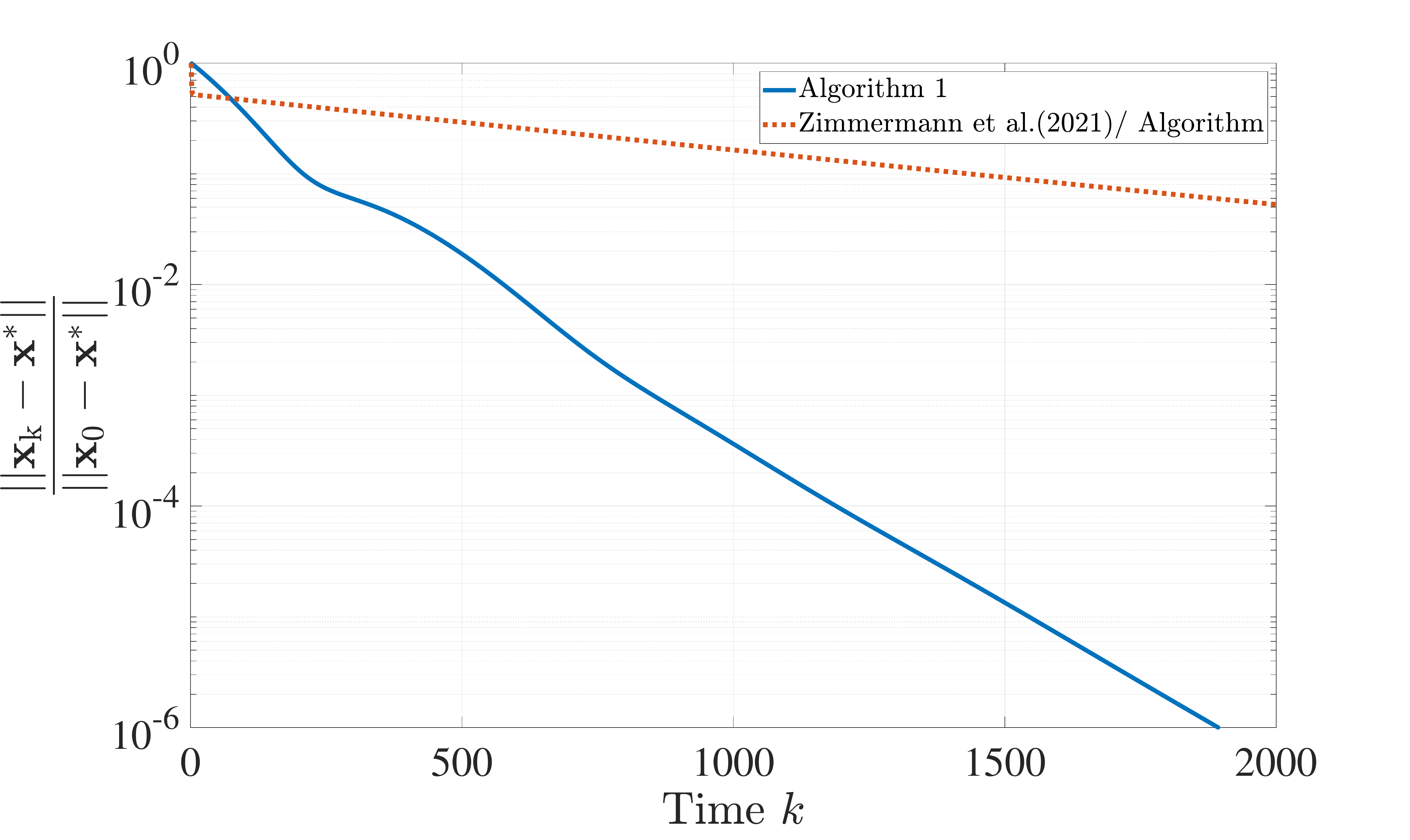}
	{Fig.3. The trajectories of  $  \frac{\| \boldsymbol{x}_k - \boldsymbol{x}^* \|}
		{\|  \boldsymbol{x}_0 - \boldsymbol{x}^*     \|}$ of Algorithm 1 and Algorithm in \cite{zimmermann2021solving} .  }
\end{figure}
\begin{figure}[htbp]
	\centering
	\includegraphics[width=3.5in]{ 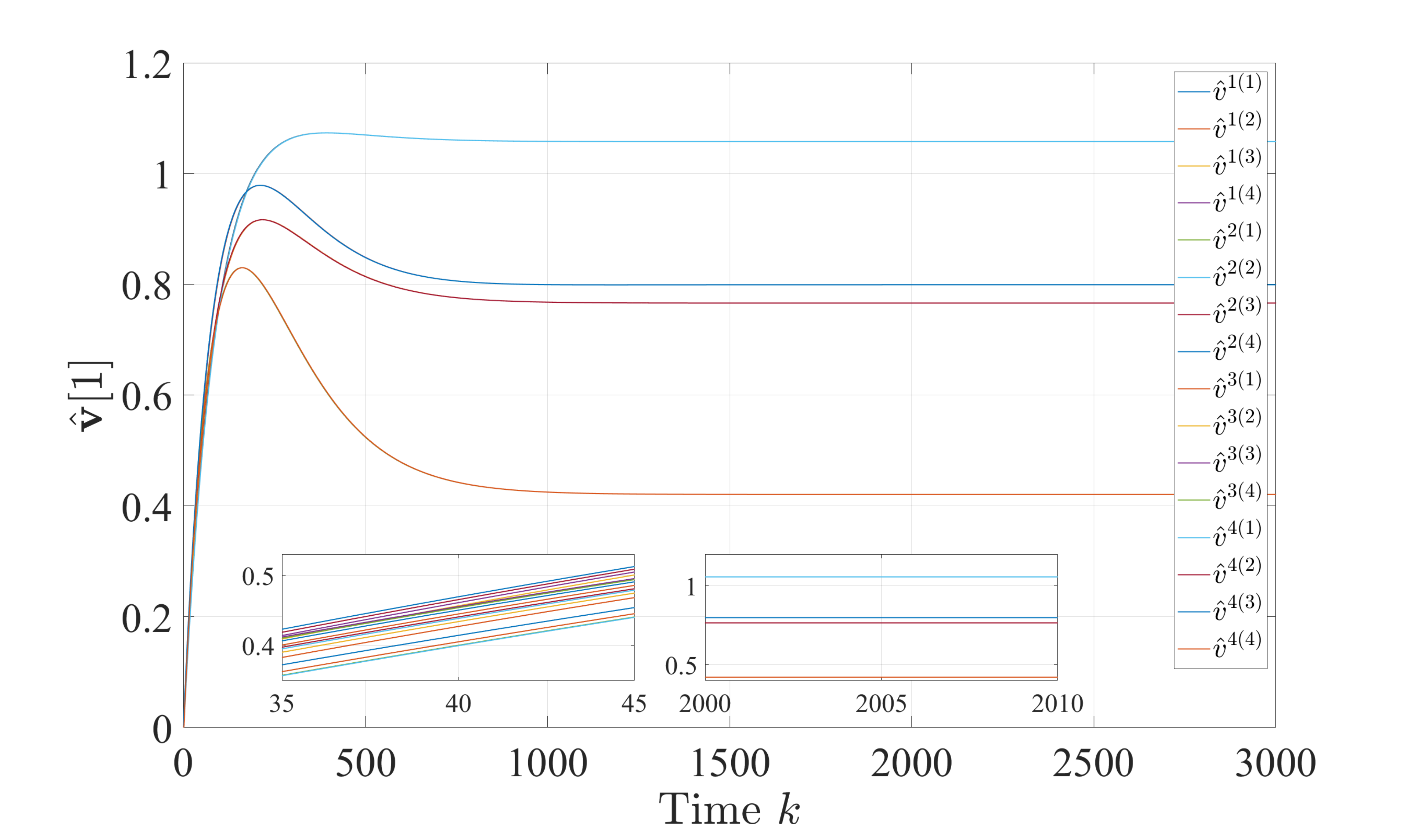}
	{Fig.4.  The  trajectories of 
		$ \hat{\boldsymbol{v}}[\mathbf{1}]  $. }
\end{figure} 

 In order to show the  communication advantages of  multi-aggregative games.
  We present the consensus process of two estimators which are $ \hat{\boldsymbol{v}} $ in this work and $ \hat{\boldsymbol{q}} $ introduced in \cite{nian2021distributed} for comparison.
  In our setting, we use 3-dimension $ x^j_i  $ to be hub's decision,  and hence the aggregate quantity $ v $ is also  a 3-dimension vector.
  Thus, the estimator $ \hat{\boldsymbol{v}} $ is a 12-dimension vector for 4 clusters.
  However, in \cite{nian2021distributed}, the estimator $ \hat{\boldsymbol{q}} $ has to be a 102-dimension vector, since it estimates all
  34 agents' decisions. 
  Then we let $ \hat{\boldsymbol{v}} $ and $ \hat{\boldsymbol{q}} $ reach consensus in $ \mathcal{G}^0 $ to $ \bar{\boldsymbol{v}} $ and $ \bar{\boldsymbol{q}} $ which are randomly drawn from $ [30,50] $ respectively.
   As is depicted in Fig.5, where $ \epsilon $ represents the upperbound of $\| \hat{\boldsymbol{v}}(\hat{\boldsymbol{q}}) - \bar{\boldsymbol{v}}(\bar{\boldsymbol{q}})   \|  $, 
   that estimator $ \hat{\boldsymbol{q}} $ cost more time compared to $ \hat{\boldsymbol{v}} $.
However, this is a conservative gap between these two kinds of estimators, since $ \hat{\boldsymbol{q}} $'s dimensions will keep expanding with the increase of the agents while $ \hat{\boldsymbol{v}} $ is not.
Additionally,  the whole system requires less mutually communications between individuals (agents, coordinators) by hierarchical scheme in this work compared to those network schemes without coordinators, such as \cite{deng2021generalized}, \cite{zimmermann2021solving}, and \cite{nian2021distributed}.
Thus, the multi-cluster aggregative game model is more suitable for large-scale participant interaction scenarios compared to conventional multi-cluster games.

\begin{figure}[htbp]
	\centering
	\includegraphics[width=3.5in]{ 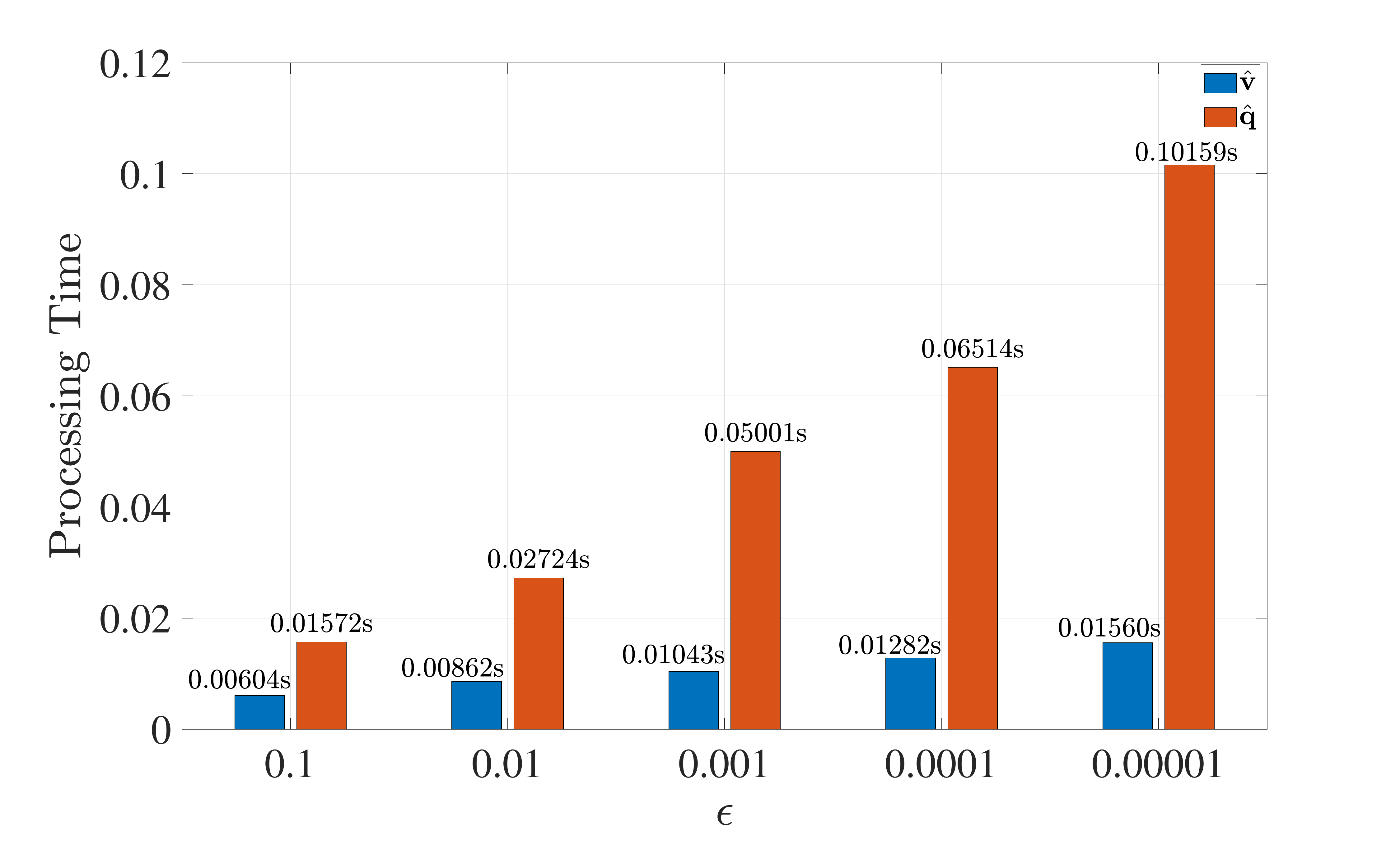}
	{\justifying Fig.5. The processing time of the convergence process of $ \hat{\boldsymbol{v}}(\hat{\boldsymbol{q}}) $ to $ \bar{\boldsymbol{v}}(\bar{\boldsymbol{q}}) $ under different $ \epsilon $. }
\end{figure}
\par In addition, we set $\textbf{a}=[1,1,1,1] $, $\tilde{\textbf{a}}=[0.9,0.6,0.5,0.1] $, termed impact factors, which implies the different policies corresponding to  different communities.
These impact factors  can be regarded as the government's energy policies for different regions.
$\textbf{a}$ means that the price is completely dependent on the aggregate quantity of the whole game
$ \frac{1}{n}\sum_{j=1}^{m}\sum_{i=1}^{n_j}x^j_i $, the same setting can be referenced in \cite{kebriaei2021multipopulation}.
Meanwhile, $ \tilde{\textbf{a}} $ states for different percentage of other communities to influence the corresponding cluster, see in subsection \ref{case2}.
Form Fig.6, we can see that the price is higher with parameter $ \tilde{\textbf{a}} $ compared to $ \textbf{a} $,  and the same with $ x^j $. It results in a higher payoff.
Namely, communities tend to generate more power with the higher energy price, and it yields a higher return.
Thus, it is a more flexible way corresponding to different energy policies in different regions.
\begin{figure}[htbp]
	\centering
		\includegraphics[width=3.5in]{ 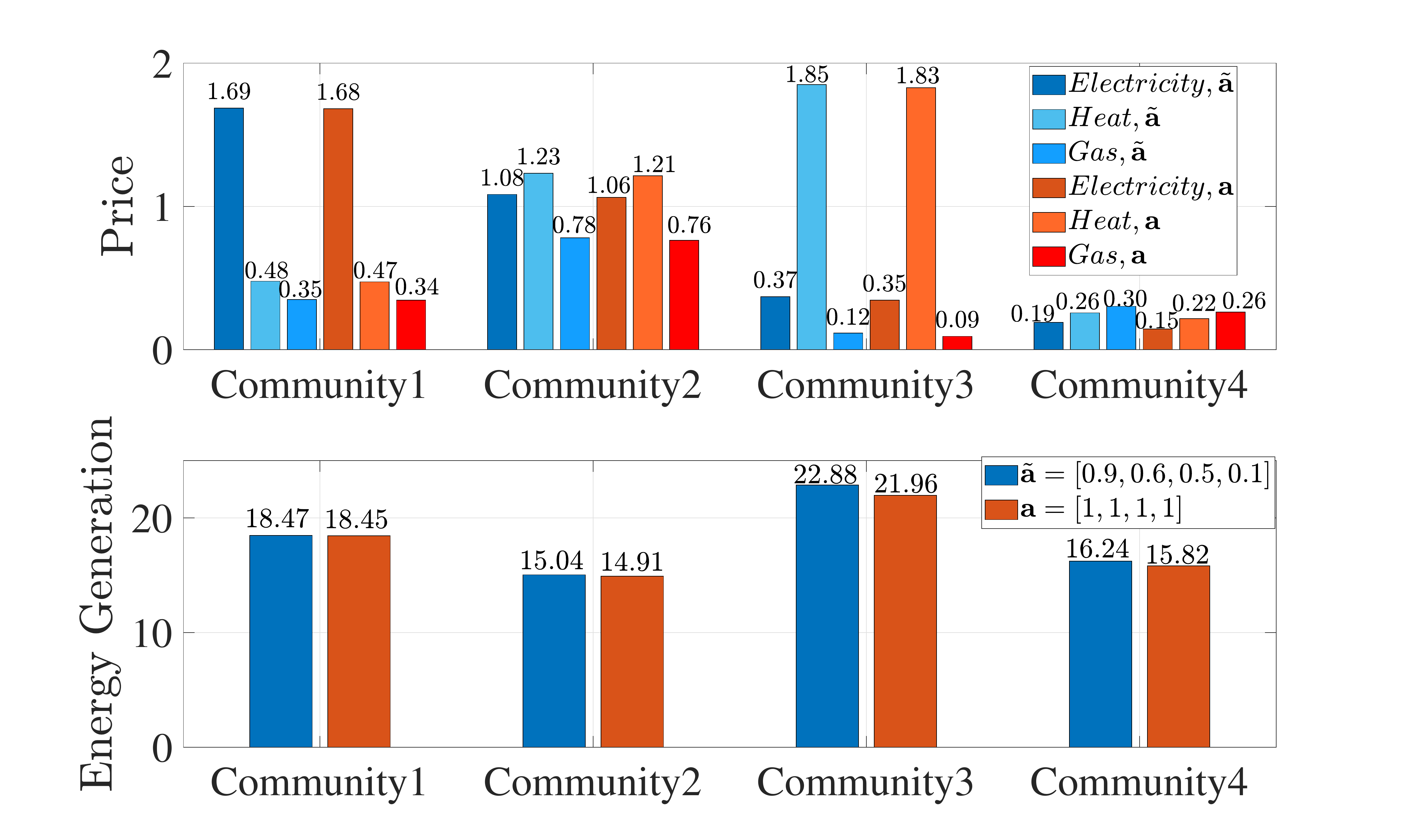}
     	\includegraphics[width=3.5in]{ 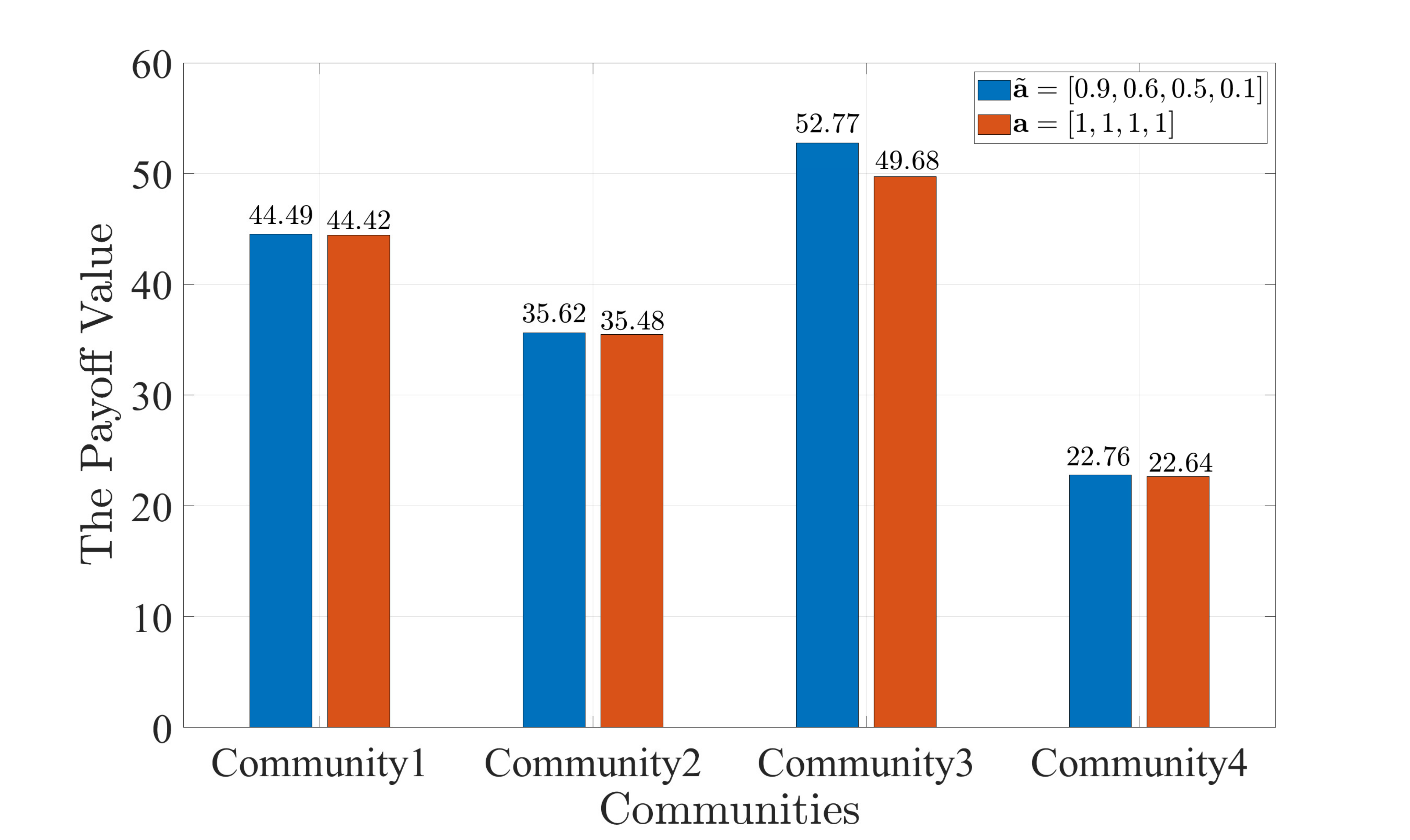}
	{\justifying Fig.6. The prices, aggregate energy generation,  and the payoff values with respect with different communities under $ \tilde{\textbf{a}} $
		and $ \textbf{a} $. }
	\end{figure}
\subsection{Case 2: Large-Scale Energy Subnets}
\par In this subsection, we show the ``anonymity'', ``lightweight'', and ``plug and play'' features by extending the scale size of each subnet and the number of subnets. 
The large-scale energy subnets diagram is shown in Fig.7.
\par To show the ``anonymity'' feature of Algorithm 1, we compare the algorithm's performances by adding a different number of agents to each subnet, i.e., 4 subnets with 7, 8, 9, and 10 agents, 4 subnets with 25, 30, 35, and 40 agents, and 4 subnets with 100, 150, 200, and 250 agents.
As depicted in Fig.8, the convergence rate of Algorithm 1 has barely changed, even with the number of agents having reached 100, 150, 200, and 250.
\begin{figure}[htbp]
\centering
\includegraphics[width=3.5in]{ 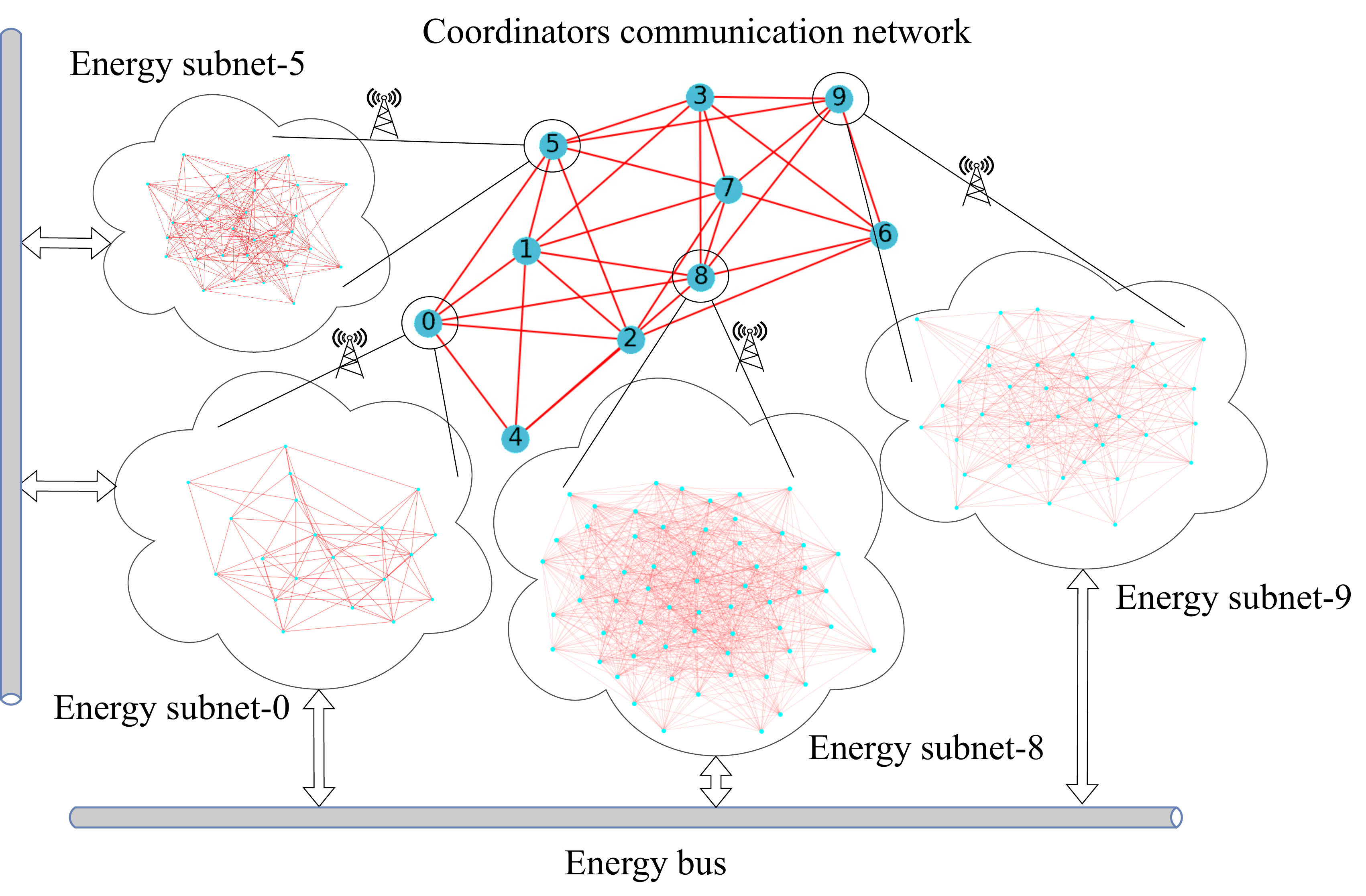}
{\justifying Fig.7. The Energy Internet system with large-scale energy subnets. }
\end{figure}
The ``anonymity'' feature of Algorithm 1 makes it more readily for extending new hubs.
\par The `` lightweight '' feature is present in the following.
Recall that $  n \triangleq\sum_{j=1}^{m} n_j $ is the total number of agents, $ m $ is the number of clusters, 
$ q $ is the strategy dimensions of each agent, 
and  we denote by $ T $  the iteration rounds of the algorithm, moreover, the date type is set to be double.
Then, as shown in table I, we can infer that for large-population clusters, i.e., $ n $ is sufficient large, the requirement calculation resource, like RAM, would cost less in Algorithm 1 than other multi-cluster algorithms.
\begin{table}[htbp] 
	\centering
	\caption{ comparison with agents' estimates of other\\  multi-cluster algorithms }
	\begin{tabular}{ccccc}
		\toprule 
		 &
		\multicolumn{2}{c}{a estimator}& \multicolumn{2}{c}{total estimators}\cr
		\cmidrule(lr){2-3}\cmidrule {4-5} 
	     	&   dimensions   & RAM(GB)  & dimensions  & RAM(GB) \cr  
	    \midrule 
	      This paper & $ q $ & $ \frac{8qT}{1024^3} $ & $ mq $ &$ \frac{8mqT}{1024^3} $\\
	    \midrule 
	       	\cite{zimmermann2021solving}, \cite{nian2021distributed} & $ nq $  & $ \frac{8nqT}{1024^3} $ & $ n^2q $ & $ \frac{8n^2qT}{1024^3} $\\
     \bottomrule
	\end{tabular}
\end{table}
For instance, in the case of 4 subnets scenario, and each subnet has 100, 150, 200, 250 agents respectively. A hub processes a estimator of 
$ 3*(100+150+200+250) $ dimensions, and the total estimators dimensions reach 
$ 3*(100+150+200+250)^2  $. 
With $ 20000 $ iterations, the algorithms require at least total 219GB RAM if there exist no simplification measurements, whereas Algorithm 1 requires only $ 0.000447 $GB. 
Thus, the ``lightweight'' feature of Algorithm 1 makes it more practical and stable in large-population cases.
\par To demonstrate the ``plug and play'' feature, we not only increased the number of agents in each subnet but also expanded the number of subnets from 4 to 6,  and 10. 
As shown in Fig.9, the performance of Algorithm 1 remains stable under different numbers of subnets and total agents, which are 4 subnets with 34 agents, 6 subnets with 415 agents, and 10 subnets with total of 1077 agents.
\par Combined with features of ``anonymity'', ``lightweight'', and ``plug and play'', Algorithm 1 is suitable for large-population multi-agent systems and can readily be implemented. 
\begin{figure}[htbp]
	\centering
	\includegraphics[width=3.5in]{ 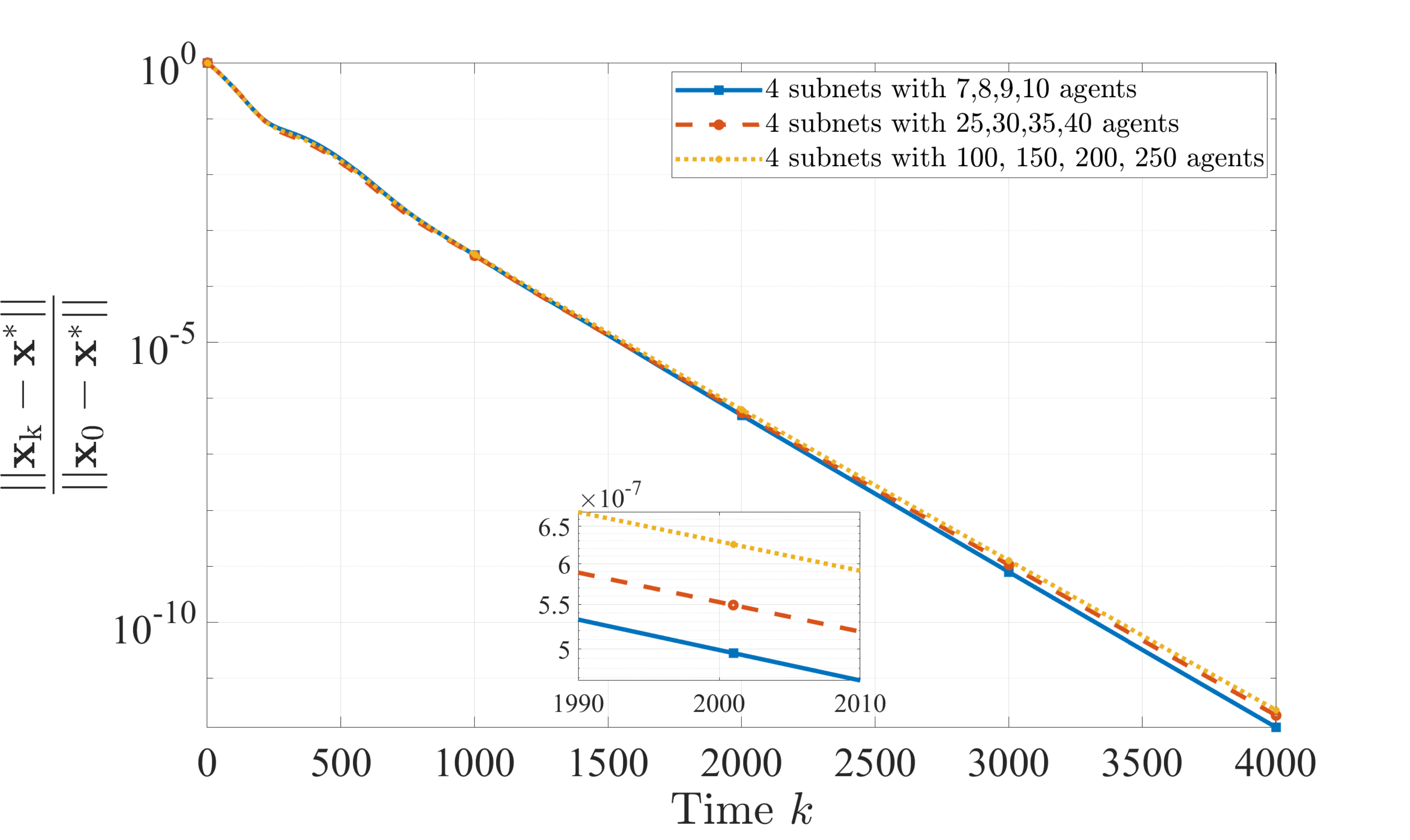}
	{\justifying Fig.8. The trajectories of  $  \frac{\| \boldsymbol{x}_k - \boldsymbol{x}^* \|}	{\|  \boldsymbol{x}_0 - \boldsymbol{x}^*     \|}$ under 4 subnets with 7, 8, 9, 10 agents, 4subnets with 25, 30, 35, 40 agents, and 4 subnets with 100, 150, 200, 250 agents. }
\end{figure}
\begin{figure}[htbp]
	\centering
	\includegraphics[width=3.5in]{ 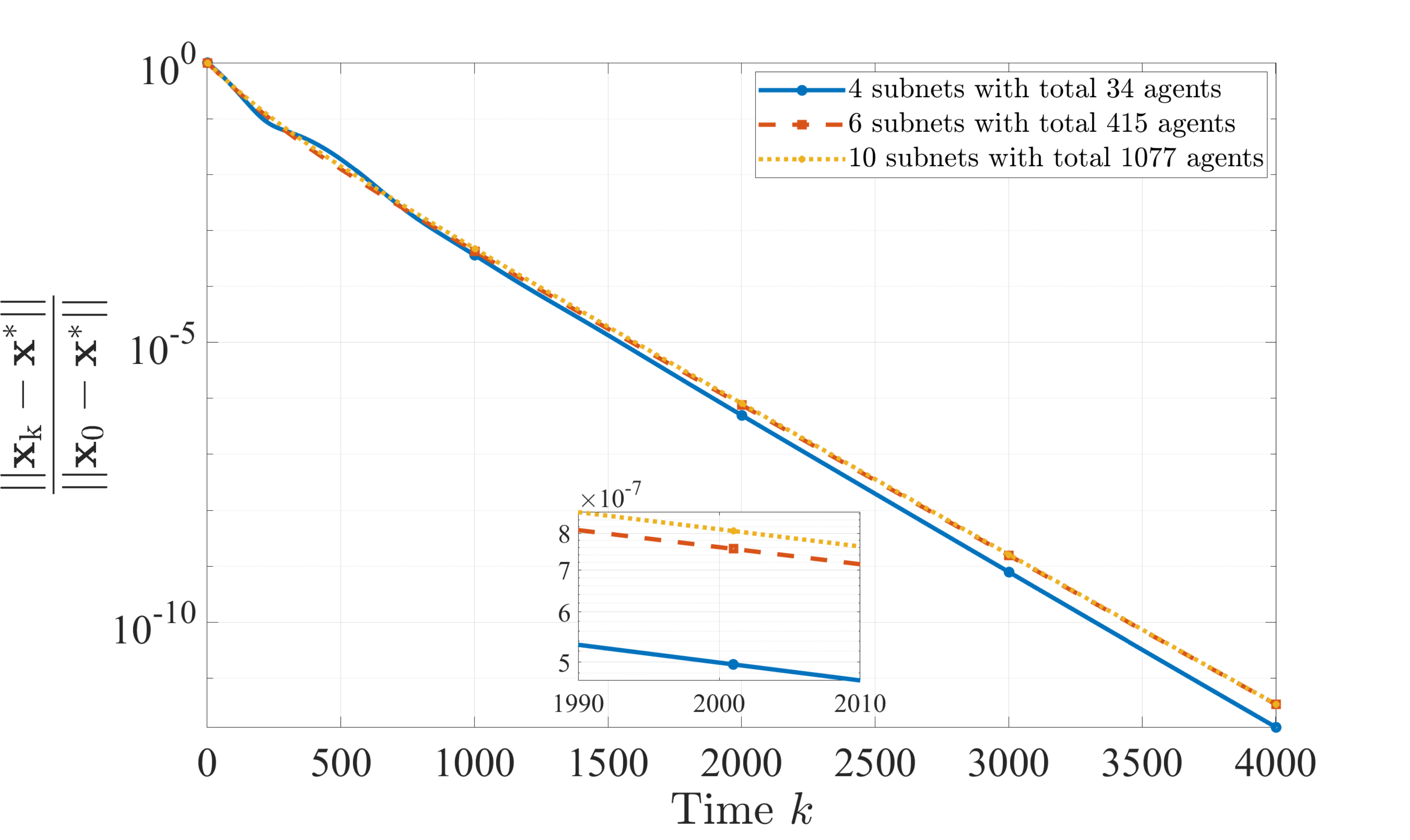}
	{\justifying Fig.9. The trajectories of  $  \frac{\| \boldsymbol{x}_k - \boldsymbol{x}^* \|}	{\|  \boldsymbol{x}_0 - \boldsymbol{x}^*     \|}$ under 4 subnets with total 34 agents, 6 subnets with total 415 agents, and 10 subnets with total 1077 agents. }
\end{figure}

\section{Conclusion}
 \label{section6}
\par In this paper, we proposed a multi-cluster aggregative game with a hierarchical communication architecture, where each cluster has a semi-decentralized structure, and coordinators are able to communicate with each other. The decisions are collected and transmitted by coordinators, while gradients required for agents are estimated through communication with their neighbors within the cluster. This framework offers the benefits of easy extension and privacy preservation.
Subsequently, we developed a linearly convergent algorithm to seek Nash equilibrium and provided a rigorous proof of convergence. By considering aggregate quantities of each cluster, multi-cluster aggregative games are particularly suitable for large-scale scenarios as they significantly reduce the communication burden between clusters. Additionally, the semi-decentralized structure of each cluster readily accommodates the ``plug and play'' needs of numerous engineering systems.
We further discussed its potential application to the Energy Internet and demonstrated its effectiveness through numerical simulations.
\section{Acknowledgment}
The authors thank Prof. Lacra Pavel for valuable comments on early versions of this work.
%

\ifCLASSOPTIONcaptionsoff
  \newpage
\fi

\bibliographystyle{IEEEtran}
\bibliography{IEEENEW}

\end{document}